\documentclass[12pt,twoside]{article}

\usepackage[T1,T2A]{fontenc}
\usepackage[utf8]{inputenc}
\usepackage[russian]{babel}

\usepackage[tbtags]{amsmath}
\usepackage{amssymb}

\DeclareMathOperator{\sgn}{\mathop{sgn}}
\DeclareMathOperator{\const}{\mathop{const}}
\DeclareMathOperator{\Sp}{\mathop{Sp}}

\usepackage{mathrsfs}
\usepackage{nicefrac}

\usepackage{cite}

\usepackage{graphicx}
\usepackage{wrapfig}

\usepackage{morefloats}

\usepackage{fullpage}

\usepackage{array}
\usepackage{tabularx}

\usepackage{indentfirst}

\selectfont

\def\cyrdash{—\penalty\@m }

\begin{document}

\title{Локализованный магнетизм в~низкоразмерных системах}

\author{А. А. Катанин и В. Ю. Ирхин}

\maketitle

Введение . . . . . . . . . . . . . . . . .  .  .  .  .  .  .  . . . . . . . . . . . . . . . . . . . . . 1

9.1.Квазидвумерные магнетики с анизотропией типа «легкая ось» . . . . 5

9.1.1. Нелинейные бозонные представления в теории квазидвумерных
ферро- и антиферромагнетиков и ССВТ квазидвумерных магнетиков (5);
9.1.2. Перенормировка вершины взаимодействия и подрешеточной намагниченности в лестничном приближении (15); 9.1.3. Теоретико-полевое
описание квазидвумерных магнетиков с локализованными моментами (19);
9.1.4. Описание различных температурных режимов в рамках ренормгруппового подхода и 1/N-разложения (22); 9.1.5. Теоретическое описание экс-
периментальных данных намагниченности и температур Нееля слоистых систем (29).

9.2.Квазидвумерные магнетики с анизотропией типа «легкая плоскость» 36

9.3.Слоистые изотропные антиферромагнетики с треугольной решеткой 41

9.4.Квазиодномерные изотропные антиферромагнетики . . . . .  .  .  . . .   . . . 43

9.4.1. Модель и самосогласованный спин-волновой подход (43); 9.4.2. Про-
цедура бозонизации (46); 9.4.3. Приближение межцепочечного среднего по-
ля для бозонизированного гамильтониана и поправки первого порядка по
$1/z_{\bot }$ (47); 9.4.4. Сравнение с экспериментальными данными (51).

Заключение . . . . . . . . . . . . . . . . . . . . . . . . . . . . . . . . . . . .  .  .  .  .  .  .51

\section*{Введение}
\addcontentsline{toc}{section}{Введение}

Исследование низкоразмерного магнетизма~— важная задача современной физики твердого тела. Экспериментальный интерес к этой проблеме связан с возможностью практических применений, например в области спинтроники и квантовых вычислений. В~настоящей главе мы ограничимся обсуждением неметаллических систем, которые хорошо описываются обычной моделью локализованных спинов Гейзенберга. 

Примеры таких систем с необычными магнитными свойствами~— слоистые перовскиты, в том числе Rb$_{2}$MnF$_{4}$, K$_{2}$NiF$_{4}$~\cite{10:001}, K$_{2}$MnF$_{4}$~\cite{10:002} (анизотропия «легкая ось»), K$_{2}$CuF$_{4}$, NiCl$_{2}$, BaNi$_{2}$(PO$_{4}$)$_{2}$~\cite{10:003} (анизотропия «легкая плоскость»), органические соединения~\cite{10:004,10:005}, ферромагнитные пленки, мультислои и поверхности~\cite{10:006}. В~конце прошлого столетия интерес к низкоразмерным соединениям возрос в связи с исследованиями магнитных свойств медь-кислородных плоскостей в высокотемпературных сверхпроводниках, в том числе на основе La$_{2}$CuO$_{4}$~\cite{10:007}.

В последнее время активно исследуются так называемые ван-дер-ваальсовы слоистые и двумерные (монослойные) системы, наиболее известными представителями которых являются CrI$_{3}$ и CrBr$_{3}$~\cite{10:008}. Магнитные свойства в таких материалах существенно зависят от структуры и количества слоев, а также чувствительны к внешним воздействиям, что обуславливает их практическую перспективность. Например, объемный кристаллический CrI$_{3}$ является ферромагнитным  с температурой Кюри $61$~К и ромбоэдрической упаковкой слоев, в то время как при малом числе слоев возникает слоистая антиферромагнитная фаза с более низкой температурой упорядочения $45$~К и моноклинной упаковкой~\cite{10:009}.

Еще один класс низкоразмерных магнитных систем с локальными моментами~— квазиодномерные соединения, содержащие цепочки магнитных атомов с маленьким межцепочечным обменом. К~ним может быть отнесено, в частности, такое хорошо экспериментально исследованное соединение как KCuF$_{3}$~\cite{10:010}, а также ряд систем на основе стронция, например Sr$_{2}$CuO$_{3}$ ($S=1/2$)~\cite{10:011,10:012} и цезия: CsNiCl$_{3}$ ($S=1$)~\cite{10:013}, CsVCl$_{3}$ ($S=3/2$)~\cite{10:014}. Родственный класс соединений представляет собой системы со «спиновыми лестницами»~— ограниченным числом цепочек магнитных атомов, связанных обменным взаимодействием~\cite{10:015}. 

В отличие от трехмерных систем, возможность магнитного упорядочения в низкоразмерных системах значительно ограничена из-за сильных флуктуаций магнитного параметра порядка. Как известно, магнитный порядок в чисто одно- и двумерных изотропных системах отсутствует при конечных температурах. Согласно теореме Мермина—Вагнера, двумерные изотропные магнетики обладают дальним порядком только в основном состоянии, а точные результаты для одномерных изотропных антиферромагнетиков свидетельствуют об отсутствии дальнего магнитного порядка даже при $T=0$. Реальные соединения обладают конечной величиной температуры магнитного перехода $T_{\text{M}} \ll |J|$ ($J$~— величина обменного взаимодействия в цепочках или в плоскости), обусловленной слабым межцепочечным (межплоскостным) обменом и~(или) анизотропией. Малость температуры перехода приводит к ряду специфических особенностей этих систем. В~частности, выше точки магнитного перехода ближний магнитный порядок полностью не разрушается (в двумерной ситуации он сохраняется до $T\sim |J|$), так что существует широкая область выше $T_{\text{M}}$ с сильным ближним порядком~\cite{10:002,10:007}. 

Существенный прогресс в понимании свойств основного состояния и термодинамики одно- и двумерных систем был достигнут благодаря  численным методам (квантовый метод Монте-Карло и метод ренормгруппы). В~то~же время такие методы не заменяют аналитических подходов, позволяющих описать термодинамические свойства слоистых систем в широком интервале температур и полезных как для теоретического понимания физических свойств этих систем, не очевидных из результатов численных расчетов, так и для практических целей описания реальных соединений. 

Стандартная теория спиновых волн~\cite{10:016,10:017,10:018} применима к низкоразмерным магнетикам лишь при низких температурах $T\ll T_{\text{M}}$. Эта теория пренебрегает взаимодействием спиновых волн, что приводит, в частности к резкому завышению температур фазового перехода низкоразмерных соединений. Проблема магнон-магнонного взаимодействия в ферромагнетиках впервые детально исследовалась в классических работах Дайсона~\cite{10:017}, построившего последовательную теорию термодинамических свойств при низких температурах. Позже эти результаты были воспроизведены Малеевым с помощью нелинейного бозонного представления спиновых операторов~\cite{10:018}. В~этом формализме проблема взаимодействия спиновых волн сводится к динамическому взаимодействию магнонов. Формализм Дайсона—Малеева был применен к проблеме взаимодействия спиновых волн в трехмерных~\cite{10:019} и двумерных~\cite{10:020} антиферромагнетиках; особое внимание в этих работах уделялось вычислению спин-волнового затухания, которое оказалось малым в широкой области импульсного пространства при достаточно низких температурах. Неаналитические поправки к спектру спиновых волн и теплоемкости низкоразмерных систем, возникающие за счет динамического взаимодействия магнонов, были исследованы в работах~\cite{10:021}. 

При температурах, не малых по сравнению с температурой магнитного перехода, существенную роль начинает играть кинематическое взаимодействие спиновых волн, возникающее вследствие ограничения числа бозонов на узле. Бозон-фермионное представление, позволяющее в явном виде учесть кинематическое взаимодействие спиновых волн, было предложено Барьяхтаром, Криворучко и Яблонским~\cite{10:022,10:023}. Введение вспомогательных фермионов в этом представлении позволяет избежать дополнительного условия для числа бозонов на узле. При не слишком низких температурах, однако, спин-волновая картина возбуждений становится полностью неадекватной и для правильного описания термодинамики необходим учет неспинволновых возбуждений. В~некоторой степени эта ситуация аналогична теории зонного магнетизма, где теория Стонера (среднего поля) неспособна адекватно описать термодинамические свойства, что стимулировало развитие спин-флуктуационных теорий~\cite{10:024}. Последние оказались особенно успешными в случае слабых зонных магнетиков, аналогичных, в некоторой степени, низкоразмерным магнитным системам с малыми значениями точки перехода. В~то время как вклад неспинволновых возбуждений в термодинамические свойства локализованных магнетиков обсуждался много лет назад в рамках феноменологической теории~\cite{10:025,10:026}, соответствующая микроскопический подход начал развиваться лишь в последнее время в рамках так называемого $1/N$-разложения~\cite{10:027}, где $N$~— число спиновых компонент ($N=3$ для модели Гейзенберга). Это разложение оказалось удивительно успешным при описании термодинамических свойств двумерных~\cite{10:027} и квазидвумерных~\cite{10:028} магнетиков.

В одномерных антиферромагнетиках картина спектра возбуждений сильно зависит от спина~$S$. Начиная с работ Бете, построившего точную волновую функцию («Бете-анзац») для одномерной антиферромагнитной цепочки, известно, что эти системы не обладают дальним магнитным порядком даже в основном состоянии. Современные теоретические подходы к одномерным системам основаны на идее Халдейна~\cite{10:030,10:031}, выполнившего преобразование проблемы цепочки к нелинейной сигма-модели. Согласно результатам Халдейна, случаи целого и полуцелого спина качественно различны. Для полуцелого спина появляется так называемый топологический член в эффективном действии, приводящий к необычному магнитному поведению таких цепочек. 

Для одной цепочки с $S=1/2$ (та~же самая ситуация имеет место при любом полуцелом значении спина), основное состояние обладает «квазидальним порядком», когда спиновые корреляции на больших расстояниях спадают по степенному, а не экспонециальному закону. Спектр возбуждений при этом является бесщелевым, хотя намагниченность равна нулю (что напоминает двумерную классическую $XY$~модель ниже точки Березинского—Костерлица—Таулеса $T_{\text{BKT}}$). В~то~же время для целых значений спина $S$ спектр возбуждений содержит так называемую халдейновскую щель порядка $\exp(-pS)$ и структура спектра возбуждений близка к предсказаниям спин-волновой теории.

В связи с «экзотическим» поведением цепочек с полуцелым спином они не могут быть исследованы в рамках спин-волновой теории и их рассмотрение требует принципиально новых физическим подходов. Для предельно квантового случая $S=1/2$ (который также наиболее важен с практической точки зрения) был развит метод бозонизации, использующий представление Йордана—Вигнера спиновых операторов через фермионные. Далее выполняется переход от фермионных операторов к бозонным, описывающим коллективные (не спин-волновые) магнитные возбуждения. Этот подход оказался также успешен при исследовании спиновых лестниц~\cite{10:015,10:032,10:033}.

Для исследования квазиодномерных систем были развиты комбинация бозонизации (и~(или) Бете-анзаца) с методом ренормгруппы~\cite{10:034,10:035,10:036} и межцепочечным приближением среднего поля~\cite{10:037}. Эти методы предсказывают конечную величину температуры магнитного перехода $T_{\text{N}} \propto |J^{\prime }|$ при сколь угодно малой величине межцепочечного взаимодействия $J^{\prime }$. В~то время как первый подход не позволяет получить каких-либо количественных оценок величины $T_{\text{N}}$, второй пренебрегает спиновыми корреляциями на разных цепочках, что приводит к резкому завышению температур Нееля по сравнению с их экспериментальными значениями. Таким образом, теория межцепочечного среднего поля приводит к тем~же трудностям при описании квазиодномерных магнетиков, что и спин-волновая теория в квазидвумерных магнетиках. Эта ситуация опять~же аналогична проблемам теории Стонера при описании зонных магнетиков.

Итак, описание квазидвумерных и квазиодномерных магнетиков требует существенно новых подходов к этим системам, рассмотрение которых и является предметом настоящей главы.

\section{Квазидвумерные магнетики с~анизотропией типа «легкая ось»}
\label{sec:10:2}

Для рассмотрения квазиодномерных и двумерных магнетиков с локализованными моментами используем модель Гейзенберга 
\begin{equation} \label{eq:10:2.1} 
\mathscr{H}=-\frac{J}{2} \sum _{i\delta _{\Vert }} \mathbf{S}_{i} \mathbf{S}_{i+\delta _{\Vert }} +\mathscr{H}_{3\textrm{D}} +\mathscr{H}_{\text{anis}} ,
\end{equation} 
\[
\mathscr{H}_{3\textrm{D}} =-\frac{J^{\prime }}{2} \sum _{i\delta _{\bot }} \mathbf{S}_{i} \mathbf{S}_{i+\delta _{\bot }} ,
\] 
\begin{equation} \label{eq:10:2.2} 
\mathscr{H}_{\text{anis}} =-\frac{J\eta }{2} \sum _{i\delta _{\Vert }} S_{i}^{z} S_{i+\delta _{\Vert }}^{z} -|J|\zeta \sum _{i} (S_{i}^{z})^{2} ,              
\end{equation} 
где $J>0$ для ферромагнетика, $J<0$ для антиферромагнетика~— обменный интеграл в плоскости, $\mathscr{H}_{3\textrm{D}}$ соответствует гамильтониану межцепочечного (межслоевого) взаимодействия, $J^{\prime }=2\alpha J$ является параметром обмена между цепочками (слоями), для определенности ниже рассматривается случай $\alpha >0$, $\delta _{\Vert }$ и $\delta _{\bot }$ обозначают ближайших соседей в пределах цепочки (слоя) и для различных цепочек (слоев). $\mathscr{H}_{\text{anis}}$~— анизотропная часть взаимодействия, возникающая в результате влияния кристаллического поля окружающих ионов; $\eta $, $\zeta >0$~— параметры обменной и одноионной анизотропии соответственно.

\subsection{Нелинейные бозонные представления в~теории квазидвумерных ферро- и антиферромагнетиков и~ССВТ квазидвумерных магнетиков}
\label{sec:10:2.1}

При достаточно низких температурах $T\ll T_{\text{M}}$ элементарными возбуждениями в магнетиках являются спиновые волны. Для описания этих возбуждений удобно перейти от спиновых операторов к бозонным. В~настоящее время используются различные представления такого вида, в частности представление Дайсона—Малеева~\cite{10:017,10:018,10:023}
\begin{equation} \label{eq:10:2.3} 
S_{i}^{+} =\sqrt{2S} b_{i},\quad S_{i}^{z} =S-b_{i}^{\dagger } b_{i} ,  
\end{equation} 
\[
S_{i}^{-} =\sqrt{2S} \left( b_{i}^{\dagger } -\frac{1}{2S} b_{i}^{\dagger } b_{i}^{\dagger } b_{i} \right) ,
\] 
($b_{i}^{\dagger }$, $b_{i}$~— магнонные бозе-операторы), которое удобно для описания магнитоупорядоченной фазы. Бозонные операторы в этом представлении должны удовлетворять условию на числа заполнения на узле $N_{bi} =\langle b_{i}^{\dagger } b_{i} \rangle <2S$, что приводит к так называемому кинематическому взаимодействию спиновых волн. Чтобы обойти эту трудность, Барьяхтар, Криворучко и Яблонский ввели представление~\cite{10:022,10:023} 
\begin{equation} \label{eq:10:2.4} 
S_{i}^{+} =\sqrt{2S} b_{i} ,\quad S_{i}^{z} =S-b_{i}^{\dagger } b_{i} -(2S+1)c_{i}^{\dagger } c_{i} ,  
\end{equation} 
\[
S_{i}^{-} =\sqrt{2S} \left(b_{i}^{\dagger } -\frac{1}{2S} b_{i}^{\dagger } b_{i}^{\dagger } b_{i} \right) -\frac{2(2S+1)}{\sqrt{2S}} b_{i}^{\dagger } c_{i}^{\dagger } c_{i} ,
\] 
содержащее помимо бозонных операторов вспомогательные псевдофермионные операторы $c_{i}^{\dagger }$, $c_{i}$, учитывающие кинематическое взаимодействие спиновых волн. В~случае антиферромагнетика с двумя подрешетками используется разбиение исходной решетки на две подрешетки, в каждой из которых используется представление \eqref{eq:10:2.4} и сопряженное ему. При низких температурах соответствующая энергия псевдофермионов порядка $|J|$, так что их вклад в термодинамические величины экспонециально мал и им можно пренебречь. В~то~же время, кинематическое взаимодействие спиновых волн становится существенным при $T\sim |J|$.

Другое полезное представление спиновых операторов~— представление швингеровских бозонов~\cite{10:038,10:039,10:040}
\begin{equation} \label{eq:10:2.5} 
\mathbf{S}_{i} =\frac{1}{2} \sum _{\sigma \sigma ^{\prime }} s_{i\sigma }^{\dagger } \boldsymbol{\sigma }_{\sigma \sigma ^{\prime }} s_{i\sigma ^{\prime }} ,
\end{equation} 
где $\boldsymbol{\sigma }$~— матрицы Паули, $\sigma , \sigma ^{\prime } =(\uparrow , \downarrow )$, так что 
\begin{equation} \label{eq:10:2.6} 
S_{i}^{z} =\frac{1}{2} (s_{i\uparrow }^{\dagger } s_{i\uparrow } -s_{i\downarrow }^{\dagger } s_{i\downarrow } ),\quad 
S_{i}^{+} =s_{i\uparrow }^{\dagger } s_{i\downarrow } ,\quad S_{i}^{-} =s_{i\downarrow }^{\dagger } s_{i\uparrow } .  
\end{equation} 
Условие 
\begin{equation} \label{eq:10:2.7} 
s_{i\uparrow }^{\dagger } s_{i\uparrow } +s_{i\downarrow }^{\dagger } s_{i\downarrow } =2S 
\end{equation} 
ограничивает число спиновых состояний и должно выполняться на каждом узле решетки. Так как одновременное изменение фаз $s_{i\uparrow }$ и $s_{i\downarrow }$ бозонов, $s_{i\sigma } \to s_{i\sigma } \exp (i\phi _{i})$ не меняет физических результатов, это представление обладает калибровочной симметрией. Этот факт может быть использован для нахождения связи представления швингеровских бозонов с известным представлением Гольштейна—Примакова~\cite{10:041}. Действительно, если  фиксировать калибровку условием эрмитовости одного из операторов $s_{i\sigma }$, например $s_{i\uparrow }$, имеем из \eqref{eq:10:2.7} 
\begin{equation} \label{eq:10:2.8} 
s_{i\uparrow } =\sqrt{2S-s_{i\downarrow }^{\dagger } s_{i\downarrow }} .  
\end{equation} 
Подставляя в \eqref{eq:10:2.6}, получаем представление Гольштейна—Примакова. Таким образом, представления швингеровских бозонов и Гольштейна—Примакова эквивалентны. Эта эквивалентность, однако, может быть нарушена в приближенных подходах. В~отличие от представления Гольштейна—Примакова (или Дайсона—Малеева), представление швингеровских бозонов может быть легко обобщено на произвольное число сортов бозонов $N\geqslant 2$, что приводит к модели с SU($N$)/SU($N-1$) симметрией и позволяет построение $1/N$-разложения~\cite{10:039}.

Взаимодействие магнонов в наинизшем (борновском) приближении рассматриваются в так называемой самосогласованной спин-волновой теории (ССВТ). Впервые эта теория была применена много лет назад к трехмерной модели Гейзенберга~\cite{10:042}; те~же самые результаты были получены позднее в рамках вариационного подхода для изотропной~\cite{10:043} и анизотропной~\cite{10:044} модели Гейзенберга. Близкие идеи использовались недавно для описания двумерных магнетиков в теории «среднего поля» для бозонных операторов~\cite{10:039,10:040,10:045}, основанной на представлении операторов спина через швингеровские бозоны, и «модифицированной спин-волновой теории»~\cite{10:046}, основанной на представлении Дайсона—Малеева. Результаты этих теорий находятся в хорошем согласии с ренормгрупповыми вычислениями~\cite{10:047,10:048} и экспериментальными данными для спектра возбуждений низкоразмерных систем~\cite{10:003}. CСВТ также применялась к квазидвумерным~\cite{10:049,10:050,10:051,10:052}, фрустрированным двумерным~\cite{10:053,10:054,10:055,10:056,10:057} и трехмерным~\cite{10:054} антиферромагнетикам. 

Для вывода уравнений ССВТ используем представление Дайсона—Малеева \eqref{eq:10:2.3}. После подстановки в гамильтониан представления спиновых операторов через бозонные, возникают члены второй и четвертой степени по бозонным операторам. В~то время как квадратичные вклады описывают распространение свободных спиновых волн, вторые соответствуют их взаимодействию. Учитывая взаимодействие спиновых волн в наинизшем приближении, т.~е. расцепляя четверные формы бозонных операторов по теореме Вика, получаем квадратичный гамильтониан ССВТ 
\begin{equation} \label{eq:10:2.9} 
\mathscr{H}=\sum _{i\delta } J_{\delta } \gamma _{\delta } (b_{i}^{\dagger } b_{i} -b_{i+\delta }^{\dagger } b_{i})-\mu \sum _{i} b_{i}^{\dagger } b_{i} ,  
\end{equation} 
где 
\begin{equation} \label{eq:10:2.10} 
\gamma _{\delta _{\bot }} =\gamma =\bar{S}+\langle b_{i}^{\dagger } b_{i+\delta _{\bot }} \rangle ,\quad 
\gamma _{\delta _{\Vert }} =\gamma ^{\prime } =\bar{S}+\langle b_{i}^{\dagger } b_{i+\delta _{\Vert }} \rangle  
\end{equation} 
— параметры ближнего порядка, определяющие спиновые корреляционные функции на соседних узлах и 
\begin{equation} \label{eq:10:2.11} 
|\langle \mathbf{S}_{i} \mathbf{S}_{i+\delta } \rangle |=\gamma _{\delta }^{2} ,  
\end{equation} 
удовлетворяющие уравнениям 
\begin{equation} \label{eq:10:2.12} 
\gamma =\bar{S}+\sum _{\mathbf{k}} N_{\mathbf{k}} \cos k_{x} ,\quad 
\gamma ^{\prime } =\bar{S}+\sum _{\mathbf{k}} N_{\mathbf{k}} \cos k_{z} .
\end{equation} 
Намагниченность ферромагнетика $\bar{S}=\langle S^{z} \rangle $ определяется полным числом бозонов:
\begin{equation} \label{eq:10:2.13} 
\bar{S}=S-\sum _{\mathbf{k}} N_{\mathbf{k}} ,
\end{equation} 
где $N_{\mathbf{k}} =N(E_{\mathbf{k}})$~— функция Бозе, причем спектр спиновых волн имеет вид 
\begin{equation} \label{eq:10:2.14} 
E_{\mathbf{q}}^{\text{SSWT}} =\Gamma _{0}^{} -\Gamma _{\mathbf{q}}^{} +\Delta -\mu ,                     
\end{equation} 
\[
\Gamma _{\mathbf{q}} =2S[\gamma |J|(\cos q_{x} +\cos q_{y} )+\gamma ^{\prime }|J^{\prime }|\cos q_{z} ],
\]
\[
\Delta =|J| \left[ (2S-1)\zeta +4\eta \frac{S^{2}}{\gamma } \right] \left( \frac{\bar{S}}{S} \right) ^{2} .
\] 
Хотя ССВТ может быть обоснована лишь при температурах $T\ll T_{\text{M}}$ (при которых существует развитый дальний порядок и взаимодействие спиновых волн мало), представляет интерес экстраполяция результатов ССВТ на более высокие температуры $T\sim T_{\text{M}}$, что позволяет сравнить результаты ССВТ с результатами более сложных теорий, рассматриваемых в разделе~\ref{sec:10:2.2}. Для продолжения теории в разупорядоченную фазу в \eqref{eq:10:2.14} введен химический потенциал бозонов $\mu $, дающий возможность удовлетворить условию ограниченности общего числа бозонов при $T>T_{\text{C}}$, где $\bar{S}=0$~\cite{10:023,10:030,10:046}. При $T<T_{\text{C}}$ имеем $\mu =0$, так что число бозонов не ограничено. Вычисление спиновых корреляционных функций показывает~\cite{10:046}, что химический потенциал непосредственно определяет корреляционную длину $\xi _{\delta }$ в направлении $\delta $ согласно соотношению 
\begin{equation} \label{eq:10:2.15} 
\xi _{\delta }^{-1} =\sqrt{-\frac{\mu }{|J_{\delta } \gamma _{\delta }|}} .  
\end{equation} 

Для антиферромагнетика уравнения ССВТ имеют вид~\cite{10:051,10:052}
\begin{equation} \label{eq:10:2.16} 
\gamma =\bar{S}+\sum _{\mathbf{k}} \frac{\Gamma _{\mathbf{k}}}{2E_{\mathbf{k}}} \cos k_{x} \coth \frac{E_{\mathbf{k}}}{2T} ,  
\end{equation} 
\[
\gamma ^{\prime } =\bar{S}+\sum _{\mathbf{k}} \frac{\Gamma _{\mathbf{k}}}{2E_{\mathbf{k}}} \cos k_{z} \coth \frac{E_{\mathbf{k}}}{2T} ,
\] 
\[
\bar{S}=S+\frac 12-\sum _{\mathbf{k}} \frac{\Gamma _{0} +\Delta -\mu }{2E_{\mathbf{k}}} \coth \frac{E_{\mathbf{k}}}{2T} ,
\] 
где 
\begin{equation} \label{eq:10:2.17} 
\gamma =\bar{S}+\langle a_{i} b_{i+\delta _{\bot }} \rangle ,\quad 
\gamma ^{\prime } =\bar{S}+\langle a_{i} b_{i+\delta _{\Vert }} \rangle  
\end{equation} 
и энергия спиновых волн равна
\begin{equation} \label{eq:10:2.18} 
E_{\mathbf{q}}^{\text{SSWT}} =\sqrt{(\Gamma _{0}^{} +\Delta -\mu )^{2} -\Gamma _{\mathbf{q}}^{2}} .
\end{equation} 
Как и для ферромагнетиков, химический потенциал бозонов $\mu $, отличный от нуля выше температуры магнитного перехода, определяет корреляционную длину согласно соотношению \eqref{eq:10:2.15}.

В основном состоянии ферромагнетика $\bar{S}_{0} =S$ и $\gamma _{0} =\gamma (T=0)=1$, подрешеточная~же намагниченность и параметр ближнего порядка двумерного антиферромагнетика отличается от этих значений из-за квантовых нулевых  колебаний спинов:  
\begin{equation} \label{eq:10:2.19} 
\bar{S}_{0} =S-\frac{1}{2} \sum _{\mathbf{k}} \left[\frac{1}{\sqrt{1-\phi _{\mathbf{k}}^{2}}} -1\right]\simeq S-0.1966,  
\end{equation} 
\begin{equation} \label{eq:10:2.20} 
\gamma _{0}=1+\frac{1}{2S} \sum _{\mathbf{k}} \left[1-\sqrt{1-\phi _{\mathbf{k}}^{2}} \right]\approx 1+\frac{0.0790}{S} ,  
\end{equation} 
где $\phi _{\mathbf{k}} =(\cos k_{x} +\cos k_{y} )/2$. При этом подрешеточная намагниченность составляет~$40\%$ от ее величины в ферромагнитном случае и совпадает с ее значением в спин-волновой теории~\cite{10:016}, перенормировка~же параметра обмена в плоскости достигает~$15\%$.  Как и в стандартной теории спиновых волн, в отсутствие анизотропии ($\delta =0$) спектр спиновых волн в упорядоченной фазе является бесщелевым и при малых~$q$ имеет вид~$E_{\mathbf{q}} =Dq^{2}$ в ФМ случае и $E_{\mathbf{q}} =cq$ в~АФМ случае, где $D$~— константа жесткости спиновых волн, $c$~— скорость спиновых волн. В~ССВТ эти параметры выражаются через параметры \eqref{eq:10:2.19}, \eqref{eq:10:2.20} согласно соотношениям 
\begin{equation} \label{eq:10:2.21} 
D=JS,\quad c=\sqrt{8} |J|\gamma S.  
\end{equation} 
Выражение для спиновой жесткости ферро- и антиферромагнетика, определенной из анализа поперечной восприимчивости, имеет вид
\begin{equation} \label{eq:10:2.22}
\rho _{\text{s}} =JS^{2}\ (\text{ФМ}), \quad  \rho _{\text{s}} =|J|\gamma S\bar{S}_{0}\ (\text{АФМ}).
\end{equation}
Перенормированные (наблюдаемые) параметры межплоскостного обмена и анизотропии, определенные из спектра возбуждений равны 
\begin{equation} \label{eq:10:2.23} 
f_{\text{r}} =\frac{\Delta }{\gamma |J|S} =\frac{1}{\gamma S} \left[ (2S-1)\zeta +4\eta \frac{S}{\gamma } \right] \left( \frac{\bar{S}}{S} \right)^{2} ,  
\end{equation} 
\begin{equation} \label{eq:10:2.24} 
\alpha _{\text{r}} =\frac{2\gamma ^{\prime }}{\gamma } =\alpha \frac{\bar{S}}{S}.  
\end{equation} 
Отметим, что в отличие от параметра внутриплоскостного обмена, перенормировка параметров $\alpha ,\eta ,\zeta $ пропорциональна намагниченности, и, таким образом, обладает сильной температурной зависимостью.

При конечных температурах в отсутствии межплоскостного обмена и анизотропии ($J^{\prime } =0$, $\delta =0$) дальний порядок отсутствует в соответствии с теоремой Мермина—Вагнера, так что $\bar{S}=0$, $\mu <0$. Этот факт является следствием расходимости интегралов в уравнениях \eqref{eq:10:2.12} и \eqref{eq:10:2.16} при $T>0$ и $\mu =0$, приводящей к отсутствию решений с ненулевой намагниченностью, При низких температурах $T\ll |J|S^{2}$ абсолютная величина химического потенциала бозонов экспоненциально мала, так что корреляционная длина $\xi =\sqrt{-|J\gamma |/\mu }$ экспоненциально велика (так называемый перенормированный классический режим),   
\begin{equation} \label{eq:10:2.25} 
\xi =C_{\xi }^{\text{F}} \sqrt{\frac{J}{T}} \exp \left( \frac{2\pi \rho _{\text{s}}}{T} \right) \quad (\text{ФМ}),  
\end{equation} 
\begin{equation} \label{eq:10:2.26} 
\xi =C_{\xi }^{\text{AF}} \frac{J}{T} \exp \left( \frac{2\pi \rho _{\text{s}}}{T} \right) \quad (\text{АФМ}),  
\end{equation} 
где $C_{\xi }^{\text{F},\text{AF}}$~— зависящие от спина константы. Результаты \eqref{eq:10:2.25}, \eqref{eq:10:2.26} согласуются с результатами однопетлевого ренормгруппового (РГ) подхода~\cite{10:047,10:048}. Двухпетлевой РГ анализ изменяет только предэкспонециальный множитель: в АФМ случае он становится температурно-независимой постоянной~\cite{10:047}, в то время как в ФМ случая пропорционален $\sqrt{T/J}$ (см.~\cite{10:048}). 

В присутствии межплоскостного обмена при не слишком высоких температур $T<T_{\text{M}}$ (температура магнитного упорядочения $T_{\text{M}}$ будет рассчитана ниже) появляется дальний магнитный порядок, при этом уравнения \eqref{eq:10:2.12} и \eqref{eq:10:2.16} имеют решения с~$\bar{S}>0$. 

При низких температурах ($T\ll |J^{\prime }|S$) и произвольном $J^{\prime } /J$ поправки к намагниченности основного состояния ферромагнетика пропорциональны $T^{3/2}$, в то время как параметры ближнего порядка имеют более слабую $T^{5/2}$~— зависимость, для антиферромагнетика соответствующие зависимости~— $T^{2}$ и $T^{4}$~\cite{10:052}. При $T>T_{\text{M}}$ снова имеем $\bar{S}=0$ и $\mu <0$, так~же как в двумерном случае при конечных~$T$.

\begin{figure}[htbp]
\centering
\includegraphics{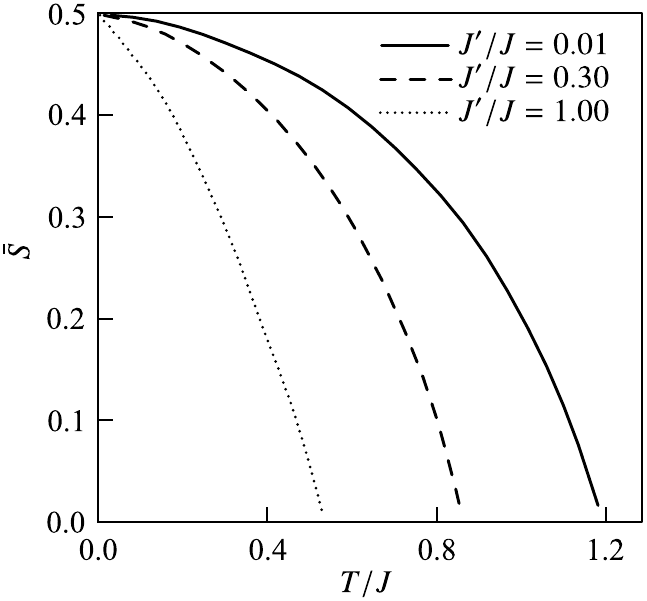}
\caption{Температурная зависимость намагниченности квазидвумерных ферромагнетиков при разных значениях отношения обменных  интегралов между плоскостями и в плоскости $J^{\prime} /J$ ($S=1/2$)}
\label{fig:10:001}
\end{figure}

\begin{figure}[htbp]
\centering
\includegraphics{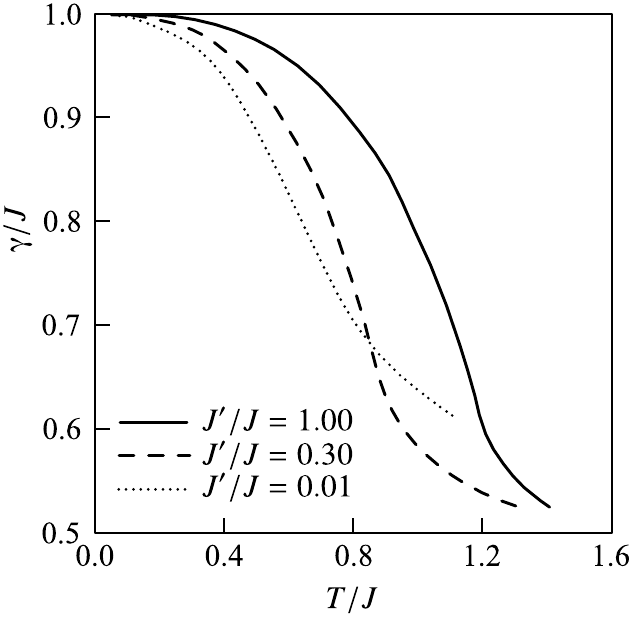}
\caption{Зависимость параметра ближнего порядка $g$ от температуры при тех~же значениях параметров, что и на рисунке~\ref{fig:10:001}}
\label{fig:10:002}
\end{figure}

\begin{figure}[htbp]
\centering
\includegraphics{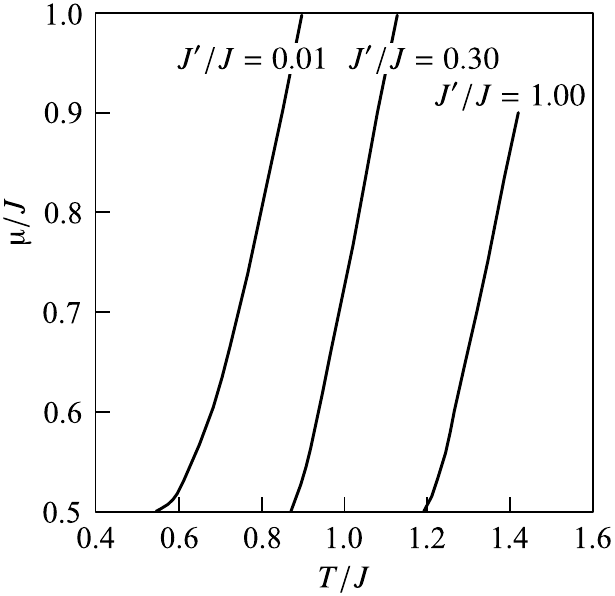}
\caption{Зависимость щели в спектре бозонов от температуры при тех~же значениях параметров, что и на рисунке~\ref{fig:10:001}}
\label{fig:10:003}
\end{figure}

Для численного исследования температурной зависимости намагниченности и параметров ближнего порядка при не слишком малых значениях межслоевого обмена удобно использовать приближение эффективного параметра ближнего порядка, производя замену~\cite{10:052}
\begin{equation} \label{eq:10:2.27} 
\sum _{\delta } J_{i,i+\delta } \gamma _{\delta } (b_{i}^{\dagger } b_{i} -b_{i}^{\dagger } b_{i+\delta } )\to \gamma _{\text{eff}} \sum _{\delta } J_{i,i+\delta } (b_{i}^{\dagger } b_{i} -b_{i}^{\dagger } b_{i+\delta } ).  
\end{equation} 
Температурная зависимость намагниченности и параметра ближнего порядка ферромагнетика для различных $J^{\prime }/J$ показаны на рисунках~\ref{fig:10:001}—\ref{fig:10:003}. При малых $T-T_{\text{M}}$ имеем $-\mu \propto (T-T_{\text{M}} )^{2}$ (см. рис.~\ref{fig:10:003} для ферромагнитного случая, та~же самая ситуация имеет место в АФМ случае), так что согласно \eqref{eq:10:2.15} критический индекс для корреляционной длины $\nu =1$. Так как намагниченность изменяется линейно около $T_{\text{M}}$, критический индекс намагниченности $\beta =1$. Влияние поправок более высокого порядка по $1/S$ на значение критических индексов обсуждается ниже. В~классическом пределе $S\to \infty $ уравнения ССВТ упрощаются и при $T<T_{\text{M}}$ ($\mu =0$) усредненный (по ближайшим соседям) параметр ближнего порядка
\begin{equation} \label{eq:10:2.28} 
\gamma _{\text{eff}} (T)=\frac{4J\gamma +2J^{\prime } \gamma ^{\prime } }{J_{0}}
\end{equation} 
(но не намагниченность!) удовлетворяет стандартному уравнению среднего поля 
\begin{equation} \label{eq:10:2.29} 
\frac{\gamma _{\text{eff}}}{S}=B_{\infty } \frac{J_{0} \gamma _{\text{eff}} S}{T}, 
\end{equation} 
где $B_{\infty } (x)=\coth x-1/x$~— функция Ланжевена (функция Бриллюэна в классическом пределе). Температура $T^{*}$, при которой $\gamma _{\text{eff}} (T^{*} )=0$, оказывается выше чем температура магнитного фазового перехода $T_{\text{M}}$, так что $\gamma _{eff} (T_{\text{M}})>0$, а поведение $\gamma _{eff}$ при $T>T_{\text{M}}$ более сложно, чем~\eqref{eq:10:2.29}. 

При малых значениях межплоскостного обмена $J^{\prime } /J\ll 1$ и анизотропии $\eta ,\zeta \ll 1$ возможно получение аналитических результатов для температурной зависимости намагниченности в широком диапазоне температур~\cite{10:049,10:052}. При этом СCВТ приводит к различным результатам для намагниченности в «квантовом» и «классическом» температурных режимах. Оказывается, что эти режимы не связаны однозначно со случаем квантовых ($S\sim 1$) и классических ($S\gg 1$) спинов (хотя классический режим реализуется лишь при $S\gg 1$), поскольку реальные критерии зависят от температуры (см. ниже). 

В квантовом режиме, который имеет место при не слишком низких температурах
\[
J^{\prime } S \ll T\ll JS \quad (\text{ФМ}),
\] 
\begin{equation} \label{eq:10:2.30} 
(JJ^{\prime })^{1/2} S \ll T\ll |J|S \quad (\text{АФМ}) 
\end{equation} 
(подрешеточная) намагниченность равна  
\begin{equation} \label{eq:10:2.31} 
\bar{S}=\begin{cases}
S-\dfrac{T}{4\pi JS}\ln\dfrac{T}{J^{\prime}\gamma^{\prime}S} & (\text{ФМ}),\\
\bar{S}_{0}-\dfrac{T}{4\pi|J|\gamma S}\ln\dfrac{T^{2}}{8JJ^{\prime}\gamma\gamma^{\prime}S^{2}}\quad & (\text{АФМ}).
\end{cases}
\end{equation}
Параметры ближнего порядка определяются соотношениями $\gamma \simeq \gamma _{0}$ и 
\begin{equation} \label{eq:10:2.32} 
\gamma^{\prime}=\begin{cases}
S-\dfrac{T}{4\pi JS}\left(\ln\dfrac{T}{J^{\prime}\gamma^{\prime}S}-1\right) & (\text{ФМ}),\\
\bar{S}_{0}-\dfrac{T}{4\pi|J|\gamma S}\left(\ln\dfrac{T^{2}}{8JJ^{\prime}\gamma\gamma^{\prime}S^{2}}-1\right)\quad & (\text{АФМ}),
\end{cases}
\end{equation}
так что $\gamma _{0}^{\prime } =\bar{S}_{0}$. Отметим, что в квантовом режиме \eqref{eq:10:2.30} интегралы по квазиимпульсам в уравнениях ССВТ определяются вкладом квазиимпульсов $q<q_{0}$, где
\begin{equation} \label{eq:10:2.33} 
q_{0}=\begin{cases}
\sqrt{T/JS}\quad & (\text{ФМ}),\\
T/c & (\text{АФМ}),
\end{cases}
\end{equation}
а не всей зоной Бриллюэна. Для критических температур в режиме \eqref{eq:10:2.30} получаем результаты 
\begin{equation} \label{eq:10:2.34} 
T_{\text{C}} =\frac{4\pi JS^{2}}{\ln (T/J^{\prime } \gamma _{\text{c}}^{\prime } S)} , 
\end{equation} 
\[
T_{\text{N}} =\frac{4\pi |J|\gamma _{\text{c}} \bar{S}_{0}}{\ln (T^{2} /8JJ^{\prime } \gamma _{\text{c}} \gamma _{\text{c}}^{\prime } S^{2} )} ,
\] 
где $\gamma _{\text{c}} \simeq \gamma _{0}$ и $\gamma _{\text{c}}^{\prime }$~— перенормированные обменные параметры в $T_{\text{M}} =T_{\text{C}}\ (T_{\text{N}})$; значение $\gamma _{\text{c}}^{\prime }$, определенное из \eqref{eq:10:2.32}, есть
\begin{equation} \label{eq:10:2.35} 
\gamma _{\text{c}}^{\prime } =(T_{\text{M}} /4\pi |J|\gamma _{\text{c}} S^{2} )J^{\prime } .                                 
\end{equation} 
Перенормировка межплоскостного обмена в \eqref{eq:10:2.34} приводит к существенному понижению температуры Кюри (Нееля) по сравнению с ее значением в спин-волновой теории, поскольку $\gamma _{\text{c}} \gamma _{\text{c}}^{\prime } /JJ^{\prime } =T_{\text{M}} /4\pi JS^{2} \ll 1$. 

В случае больших $S$ (классический предел) получаем для ферро- и антиферромагнетиков при $T\gg |J|S$
\begin{equation} \label{eq:10:2.36} 
\bar{S}=S-\frac{T}{4\pi |J|S} \ln \frac{q_{0}^{2} J}{J^{\prime } \gamma ^{\prime }} ,  
\end{equation} 
\begin{equation} \label{eq:10:2.37} 
\gamma ^{\prime } =S-\frac{T}{4\pi |J|S} \left(\ln \frac{q_{0}^{2} J}{J^{\prime } \gamma ^{\prime }} -1\right).  
\end{equation} 
В отличие от квантового случая, результаты для намагниченности в этом пределе неуниверсальны, т.~к. зависят от типа решетки через параметр обрезки $q_{0}^{2}$ (для квадратной решетки $q_{0}^{2} =32$).  Соответствующее выражение для критической температуры классического магнетика с  $1\ll \ln (q_{0}^{2} J/J^{\prime } )\ll 2\pi S$ имеет вид
\begin{equation} \label{eq:10:2.38} 
T_{\text{M}} =\frac{4\pi |J|S^{2}}{\ln (q_{0}^{2} J/J^{\prime } \gamma _{\text{c}}^{\prime } )} ,  
\end{equation} 
где $\gamma _{\text{c}}^{\prime } =T_{\text{M}} /4\pi |J|S$. Как и должно быть, критическая температура одинакова для классических ферро- и антиферромагнетиков. С~логарифмической точностью в этом случае воспроизводятся результаты спин-волновой теории, где $\gamma _{\text{c}}^{\prime } /S\to 1$. Аналогичные результаты могут быть получены в случае малой анизотропии «легкая ось»~\cite{10:029,10:066} 
\begin{equation} \label{eq:10:2.39} 
T_{\text{C}} =\frac{4\pi JS^{2}}{\ln (T/\Delta _{\text{c}} )} ,  
\end{equation} 
\[
T_{\text{N}} =\frac{4\pi |J|S\bar{S}_{0} \gamma _{\text{c}}}{\ln (T^{2} /8J\gamma _{\text{c}} S\Delta _{\text{c}} )} .
\] 
где величина щели в спектре спиновых волн $\Delta _{\text{c}} =\Delta (T_{\text{M}} )$ не может быть определена в рамках спин-волновой теории. В~пределе больших $S$ находим как для ферро-, так и для антиферромагнетиков 
\begin{equation} \label{eq:10:2.40} 
T_{\text{M}} =\frac{4\pi |J|S^{2}}{\ln (|J|Sq_{0}^{2} /\Delta _{\text{c}} )} . 
\end{equation} 

Результаты \eqref{eq:10:2.34} и \eqref{eq:10:2.38} могут быть сопоставлены с результатом приближения Тябликова. В~этом приближении (в случае ферромагнетика) интерполяционные выражения для запаздывающей коммутаторной спиновой функции Грина и перенормированного спектра спиновых волн имеют вид
\begin{equation} \label{eq:10:2.41} 
\langle \!\langle S_{\mathbf{q}}^{+} |S_{-\mathbf{q}}^{-} \rangle \!\rangle _{\omega } =\frac{2\langle S^{z} \rangle }{\omega -\omega _{\mathbf{q}}} ,\quad 
\omega _{\mathbf{q}} =2\langle S^{z} \rangle (J_{0} -J_{\mathbf{q}} )+h 
\end{equation} 
($h$~— магнитное поле). Метод Тябликова грубо описывает термодинамику обычно трехмерной модели Гейзенберга как при высоких, так и при низких температурах, хотя члены более высокого порядка в низкотемпературном разложении не вполне согласуются с результатами Дайсона~\cite{10:015} из-за неправильного учета взаимодействия спиновых волн (в частности, появляются «лишние» $T^{3}$-члены в намагниченности). Многочисленные попытки улучшить область спин-волнового описания, используя более сложные процедуры расцепления (см.~\cite{10:058}), привели к ухудшению интерполяции. Вблизи $T_{\text{M}}$ теория Тябликова дает такое~же поведение намагниченности (подрешетки), как и теория среднего поля.

Значение температуры Кюри в приближении Тябликова для произвольного значения $S$ равно
\begin{equation} \label{eq:10:2.42} 
T_{\text{C}} =\frac{S(S+1)}{3} \left(\sum _{\mathbf{q}} \frac{1}{J_{0} -J_{\mathbf{q}}} \right)^{-1} .  
\end{equation} 
Соответствующее выражение для антиферромагнетика получается заменой  $J_{0} \to J_{\mathbf{Q}}$. Для температуры магнитного перехода слоистых соединений интегрирование дает~\cite{10:028}
\begin{equation} \label{eq:10:2.43} 
T_{\text{M}} \simeq \frac{4\pi |J|S(S+1)}{3\ln (|J|q_{0}^{2} /J^{\prime } )} 
\end{equation} 
с $q_{0}^{2} =32$. Результат \eqref{eq:10:2.43} для спина~$1/2$ численно меньше, чем значение ССВТ \eqref{eq:10:2.34}, а потому лучше описывает экспериментальные данные (см. раздел~\ref{sec:10:2.5}). С~другой стороны, в классическом пределе $S\to \infty $, \eqref{eq:10:2.42} совпадает с результатом для сферической модели~\cite{10:067}, что подтверждает его интерполяционный характер. В~связи со сложностью улучшения приближения Тябликова, оно может быть частично удовлетворительно с практической, но не с теоретической точки зрения.

Хотя с логарифмической точностью все обсуждавшиеся подходы приводят в квантовом пределе $S=1/2$ к одному и тому~же значению температуры Нееля, эта точность недостаточна для количественного описания экспериментальных данных, критическое поведение описывается спин-волновыми теориями также неправильно. Формально ССВТ соответствует пределу $N\to \infty $ в SU($N$)/SU($N-1$) обобщении модели Гейзенберга~\cite{10:039}. Чтобы улучшить описание критической области и вычисление температур Кюри (Нееля), необходимо рассмотреть флуктуационные поправки к результатам теории спиновых волн более аккуратно, чем в ССВТ. Вычисление поправок первого порядка по~$1/N$ в~SU($N$)/SU($N-1$) модели может позволить описать область низких и промежуточных температур $T\lesssim T_{\text{M}}$, но неспособно правильно описать критическое поведение~\cite{10:059}. Проблемы этого подхода в критической области связаны с тем, что в указанном обобщении модели Гейзенберга  возбуждения неспинволнового  характера представляются как связанные состояния спиновых волн~\cite{10:060} и их рассмотрение весьма затруднительно в рамках $1/N$-разложения. В~связи с этим, необходимо  развитие подходов, позволяющих описать как область промежуточных температур, так и критическую область. Такие подходы рассматриваются ниже.

\subsection{Перенормировка вершины взаимодействия и~подрешеточной намагниченности в~лестничном приближении}
\label{sec:10:2.2}

\begin{figure}[htbp]
\centering
\includegraphics{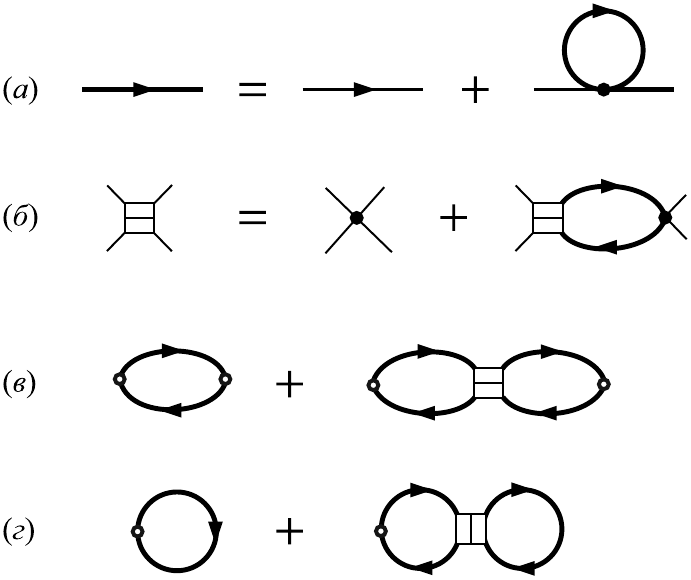}
\caption{Диаграммное изображение: (\textit{а})~хартриевских поправок к спектру магнонов, рассматриваемых в самосогласованной спин-волновой теории;  (\textit{б})~лестничного уравнения для вершины взаимодействия магнонов; (\textit{в})~поправок к (подрешеточной) восприимчивости; (\textit{г})~поправок к (подрешеточной) намагниченности}
\label{fig:10:004}
\end{figure}

В данном подразделе рассмотрим поправки к ССВТ, определяемые диаграммами второго и более высокого порядка по~$1/S$. Указанные поправки можно разделить на собственно-энергетические (приводящие к перенормировке энергии одночастичных возбуждений и возникновению их затухания) и поправки к вершине магнон-магнонного взаимодействия. Как обсуждалось выше, ССВТ удовлетворительно описывает спектр возбуждений (этот спектр уже перенормирован в соответствии с диаграммами первого порядка, см. рис.~\ref{fig:10:004}\textit{а}). Вычисление затухания спиновых волн, возникающего во втором и более высоких порядках теории возмущений, показывает, что оно относительно мало в широкой температурной области~\cite{10:020,10:061}.

Таким образом, вершинные поправки представляются наиболее существенными в двумерных системах. Для их вычисления рассмотрим сначала суммирование лестничных диаграмм для вершины на примере двумерных магнетиков с анизотропией типа «легкая ось». Как следует из РГ рассмотрения следующего подраздела, лестничное приближение дает правильный ответ для ведущей поправки к намагниченности ССВТ, полученной суммированием ведущих и субведущих сингулярных вкладов во всех порядках теории возмущений. Это связано с сокращением нелестничных вкладов в более общем паркетном приближении, см. подробное обсуждение в~\cite{10:062}.

Для формулировки лестничного приближения запишем гамильтониан ферромагнетика в виде
\begin{equation} \label{eq:10:2.44} 
\mathscr{H}=\sum _{\mathbf{q}} E_{\mathbf{q}}^{\text{SSWT}} b_{\mathbf{q}}^{\dagger } b_{\mathbf{q}} +\frac{1}{4} \sum _{\mathbf{q}_{1} \ldots \mathbf{q}_{4}} \phi (\mathbf{q}_{1}, \mathbf{q}_{2} ;\mathbf{q}_{3}, \mathbf{q}_{4}) (b_{\mathbf{q}_{1}}^{\dagger } b_{\mathbf{q}_{2}}^{\dagger } b_{\mathbf{q}_{3}} b_{\mathbf{q}_{4}} )_{\text{1PI}} \delta _{\mathbf{q}_{1}+\mathbf{q}_{2}, \mathbf{q}_{3}+\mathbf{q}_{4}} ,
\end{equation} 
где 
\begin{equation} \label{eq:10:2.45} 
\phi (\mathbf{q}_{1}, \mathbf{q}_{2}; \mathbf{q}_{3}, \mathbf{q}_{4} ) =J_{\mathbf{q}_{3}}+J_{\mathbf{q}_{4}}-(1+\eta )(J_{\mathbf{q}_{1}-\mathbf{q}_{3}} -J_{\mathbf{q}_{1}-\mathbf{q}_{4}})-4J\varsigma \simeq -2J(\mathbf{q}_{1} \mathbf{q}_{2} +f) 
\end{equation} 
— вершина магнонного взаимодействия, 
\begin{equation} \label{eq:10:2.46} 
J_{\mathbf{q}} =J[2(\cos q_{x} +\cos q_{y} )+\alpha \cos q_{z} ] 
\end{equation} 
— фурье-образ обменного взаимодействия, $f=2\zeta +4\eta $, символ 1PI означает, что из четверной формы в \eqref{eq:10:2.44} исключены все парные свертки, уже учтенные самомогласованной спин-волновой теорией. Интегральное уравнение для перенормированной вершины $\Phi (\mathbf{k}, \mathbf{p}-\mathbf{q}, \mathbf{k}-\mathbf{q}; \mathbf{p})$ в лестничном приближении имеет вид~\cite{10:052}
\begin{multline} \label{eq:10:2.47} 
\Phi (\mathbf{k}, \mathbf{p}-\mathbf{q}, \mathbf{k}-\mathbf{q}; \mathbf{p})=\phi (\mathbf{k}, \mathbf{p}-\mathbf{q}; \mathbf{k}-\mathbf{q}, \mathbf{p}) - {} \\
{} - \frac{T}{(JS)^{2}} \sum _{\mathbf{s}} \frac{\phi (\mathbf{k}, \mathbf{s}-\mathbf{q}; \mathbf{k}-\mathbf{q}, \mathbf{s})}{(s^{2} +f)[(\mathbf{s}-\mathbf{q})^{2} +f]} \Phi (\mathbf{s}, \mathbf{p}-\mathbf{q}; \mathbf{s}-\mathbf{q}, \mathbf{p}) 
\end{multline} 
(мы сохраняем здесь только вклад членов с нулевой мацубаровской частотой $\omega _{n}=0$, приводящих к логарифмически расходящимся вкладам, с одновременным обрезанием интегралов по квазиимпульсам на волновом векторе $q_{0}$, определенным в \eqref{eq:10:2.33}). Результат решения уравнения \eqref{eq:10:2.47} имеет вид 
\begin{multline} \label{eq:10:2.48} 
\Phi (\mathbf{k},\mathbf{p}-\mathbf{q}; \mathbf{k}-\mathbf{q}, \mathbf{p})=\frac{2J\mathbf{k}(\mathbf{q}-\mathbf{p})}{1-(T/2\pi JS^{2} )\ln (q_{0} /\max (f^{1/2} ,q)]} +O(Jf) \simeq {} \\ 
{} \simeq \frac{2J\mathbf{k}(\mathbf{q}-\mathbf{p})}{\bar{S}/S+(Jq^{2} /2)\chi _{\mathbf{q}0}^{zz}} +O(Jf),
\end{multline}
где 
\begin{equation} \label{eq:10:2.49} 
\chi _{\mathbf{q}0}^{zz} =\frac{T}{(JS)^{2}} \sum _{\mathbf{p}} \frac{1}{(p^{2} +f)[(\mathbf{p}-\mathbf{q})^{2} +f]}
\simeq\begin{cases}
T/[2\pi(JqS)^{2}]\ln(q^{2}/f),\quad & q^{2}\gg f,\\
T/[4\pi(JS)^{2}f], & q^{2}\ll f
\end{cases}
\end{equation} 
есть продольная восприимчивость в спин-волновой теории. Как следует из результата \eqref{eq:10:2.48}, рассматриваемая вершина магнон-магнонного взаимодействия усиливается флуктуациями, как и в  стандартном приближении случайных фаз для зонных магнетиков. Результат, аналогичный \eqref{eq:10:2.48} с заменами $\bar{S}/S\to \bar{S}/\bar{S}_{0}$, $J\to |J|\gamma $, $f\to f_{\text{r}}$ может быть также получен для антиферромагнетиков.

Для статической (подрешеточной) неоднородной продольной восприимчивости (со сдвигом $\mathbf{q}\to \mathbf{q}+\mathbf{Q}$ в АФМ случае) получаем с учетом диаграмм рисунка~\ref{fig:10:004}\textit{в} результат 
\begin{equation} \label{eq:10:2.50} 
\chi _{\mathbf{q}}^{zz} =\frac{(\bar{S}/S)\chi _{\mathbf{q}0}^{zz}}{\bar{S}/S+(T/2\pi JS)\ln [\max (f_{}^{1/2} ,q)/f_{}^{1/2} ]}
=\frac{\chi _{\mathbf{q}0}^{zz}}{1+(|J|\gamma /2\bar{S})q^{2} \chi _{\mathbf{q}0}^{zz}} .
\end{equation} 
Как следует из результата \eqref{eq:10:2.50}, продольная восприимчивость имеет различную импульсную зависимость на малых и достаточно больших импульсах: 
\begin{equation} \label{eq:10:2.51} 
\chi_{\mathbf{q}}^{zz}\simeq\begin{cases}
\chi_{\mathbf{q}0}^{zz}, & Jq^{2}\chi_{\mathbf{q}0}^{zz}\ll 2\bar{S},\\
2\bar{S}/(Jq^{2}),\quad & Jq^{2}\chi_{\mathbf{q}0}^{zz}\gg 2\bar{S}.
\end{cases}
\end{equation} 
Первая строка соответствует стандартному спин-волновому результату \eqref{eq:10:2.49}, в то время как вторая описывает вклад неспинволновых степеней свободы: пренебрегая анизотропией при $q^{2} \gg f$, находим $\chi _{\mathbf{q}}^{zz} \propto 1/q^{2}$, что отвечает критическим спиновым флуктуациям и согласуется с результатом сферической модели~\cite{10:011}. 

Рассмотрим теперь поправки к намагниченности, обусловленные рассмотренными выше продольными возбуждениями. Вычисление поправок к  (подрешеточной) намагниченности $\bar{\sigma }\equiv \bar{S}/\bar{S}_{0}$ (см. диаграммы рис.~\ref{fig:10:004}\textit{г}) дает 
\begin{equation} \label{eq:10:2.52} 
\bar{\sigma }=1-\frac{T}{JS^{2}} \sum _{\mathbf{k}} \frac{1}{k^{2} +f} +\frac{T^{2}}{J^{3} S^{4}} \sum _{\mathbf{k}\mathbf{q}} \frac{\Phi (\mathbf{k}, \mathbf{k}-\mathbf{q}; \mathbf{k}-\mathbf{q}, \mathbf{k})}{(k^{2} +f)^{2} [(\mathbf{k}-\mathbf{q})^{2} +f]} .
\end{equation} 
Интегрирование в \eqref{eq:10:2.52} приводит к результату
\begin{equation} \label{eq:10:2.53} 
\bar{\sigma }=1-\frac{t}{2} \left[\ln \frac{q_{0}^{2}}{f_{\text{r}} (T)} +2\ln \frac{1}{\max (\bar{\sigma },t)} +\Phi _{\text{a}} (t/\bar{\sigma })\right] 
\end{equation} 
где $t=T/(2\pi JS)$. Функция $\Phi _{\text{a}}$ учитывает вклад несингулярных членов. Температурная перенормировка $f_{\text{r}} (T)\propto f\bar{\sigma }^{2}$, определяемая уравнением \eqref{eq:10:2.23}, увеличивает в два раза множитель перед вторым слагаемым в квадратных скобках в промежуточной температурной области. 

Как видно из результата \eqref{eq:10:2.53}, флуктуационные поправки к намагниченности уменьшают ее значение. В~зависимости от величины температуры возможны три случая: 

(а) низкие температуры, $T\ll T_{\text{M}} \sim 2\pi |J|S^{2} /\ln (q_{0}^{2} /f)$. Тогда второе условие в \eqref{eq:10:2.51} не может быть удовлетворено, и таким образом возбуждения во всей зоне Бриллюэна имеют спин-волновой характер. При этом важен только первый член в квадратных скобках \eqref{eq:10:2.53} и температурная зависимость намагниченности описывается спин-волновой теорией; 

(б) промежуточные температуры $T\sim T_{\text{M}}$, для которых выполнено условие  $(\bar{S}/S)/\ln (q_{0}^{2} /f)\ll T/2\pi |J|S^{2} \ll \bar{S}/S$. Тогда при достаточно малых $q$ имеем $\chi _{\mathbf{q}}^{zz} \simeq \chi _{\mathbf{q}0}^{zz}$, но второе условие в \eqref{eq:10:2.51} выполнено для достаточно больших $q$, так что возбуждения на соответствующих волновых векторах являются неспинволновыми (т. е. соответствуют критическим флуктуациям). Температурная зависимость намагниченности $\bar{S}(T)$ существенно модифицируется поправками к спин-волновой теории; 

(в) критическая область, $T/2\pi |J|S^{2} \gg \bar{S}/S$ ($1-T/T_{\text{M}} \ll 1$). В~этом режиме первое условие в \eqref{eq:10:2.51} удовлетворено только для~$q^{2} \ll f$, тогда как для всех остальных~$q$ возбуждения имеют неспинволновой характер. При этом вклады $\Phi _{\text{a}}$ имеют тот~же самый порядок, что и другие члены в квадратных скобках, и температурная зависимость намагниченности должна рассматриваться в рамках более сложных подходов (см. разделы~\ref{sec:10:2.3}—\ref{sec:10:2.5}). 

Для более детального анализа указанных результатов в последующих подразделах рассматривается теоретико-полевой подход к низкоразмерным магнетикам.

\subsection{Теоретико-полевое описание квазидвумерных магнетиков с~локализованными моментами}
\label{sec:10:2.3}

Для правильного описания термодинамических свойств в широком интервале температур необходимо суммирование ведущих вкладов в термодинамические величины во всех порядках теории возмущений по магнон-магнонному взаимодействию. Такой учет может быть произведен в рамках нелинейной сигма-модели. Для вывода выражений для производящего функционала используется представление когерентных состояний ${|\mathbf{n}_{i}\rangle} =\exp (-i\varphi _{i} S_{i}^{z} )\exp (-i\theta _{i} S_{i}^{y} ){|0\rangle}$~\cite{10:068,10:069}, параметризуемых векторами~$\mathbf{n}_{i}$ единичной длины с полярными координатами $(q_{i}, j_{i})$, определенных для каждого узла решетки $i$, $|0\rangle $~— собственное состояние оператора $S_{i}^{z}$ с максимальной проекцией спина: $S_{i}^{z} |0\rangle =S |0\rangle $. Преимущество использования состояний~$|\mathbf{n}_{i}\rangle $ состоит в том, что среднее значение операторов спина по ним имеет простой вид:
\begin{equation} \label{eq:10:2.54} 
\langle \mathbf{n}_{i} | S_{i}^{m} |\mathbf{n}_{i} \rangle =S n_{i}^{m} 
\end{equation} 
т.~е. когерентные состояния являются «квазиклассическими» спиновыми состояниями. Можно показать, что с помощью когерентных состояний \eqref{eq:10:2.54} производящий функционал может быть записан в виде 
\begin{equation} \label{eq:10:2.55} 
Z=\int D \mathbf{n} \exp \left\{\int \limits_{0}^{1/T} d\tau \left[ \mathbf{A} (\mathbf{n}_{i}) \frac{\partial \mathbf{n}_{i}}{\partial \tau } -\langle \mathbf{n} | \mathscr{H} |\mathbf{n}\rangle \right] \right\} ,
\end{equation} 
где первый член в показателе экспоненты учитывает динамику спинов, связанную с их квантовым характером (так называемая фаза Берри~\cite{10:030}), а второй член описывает взаимодействие спинов; интегрирование в \eqref{eq:10:2.55} производится по угловым переменным вектора $\mathbf{n}_{i}$ на каждом узле и для каждого мнимого времени $\tau $, $\mathbf{A}(\mathbf{n})$~— векторный потенциал единичного магнитного монополя, удовлетворяющий соотношению $\nabla \times \mathbf{A}(\mathbf{n})\cdot \mathbf{n}=1$.

Среднее по когерентным состояниям гамильтониана \eqref{eq:10:2.1} может быть легко вычислено с учетом соотношений \eqref{eq:10:2.54} и приводит к выражению для производящего функционала в виде 
\begin{multline} \label{eq:10:2.56} 
Z[h]=\int D \mathbf{n} \, D\lambda \exp \left\{\frac{JS^{2}}{2} \int \limits_{0}^{1/T} d\tau \sum _{i,\delta _{\Vert } ,\delta _{\bot }} \left[\frac{2 \mathrm{i}}{JS} \mathbf{A} (\mathbf{n}_{i}) \frac{\partial \mathbf{n}_{i}}{\partial \tau }+ \mathbf{n}_{i} \mathbf{n}_{i+\delta _{\Vert }} + {} \right. \right. \\
\left. \left. {} +\frac{\alpha }{2} \mathbf{n}_{i} \mathbf{n}_{i+\delta _{\bot }} +\eta n_{i}^{z} n_{i+\delta _{\Vert }}^{z} +\sgn (J)\tilde{\zeta }(n_{i}^{z})^{2} +hn_{i}^{z} +\mathrm{i}\lambda _{i} (\mathbf{n}_{i}^{2} -1)
\vphantom{\frac{2 i}{JS}}
\right]
\vphantom{\int \limits_{0}^{1/T}}
\right\} ,
\end{multline} 
где члены в экспоненте, следующие за фазой Берри последовательно отвечают обмену в плоскости, между плоскостями, двухионной и одноионной магнитной анизотропии и неоднородному внешнему полю. Последний член в экспоненте возникает вследствие ограничения $\mathbf{n}^{2}=1$. Функционал \eqref{eq:10:2.56} содержит две переменные с размерностью длины: 
\begin{equation} \label{eq:10:2.57} 
\xi _{J^{\prime }} =a/\max (\alpha ,\tilde{\zeta },\eta )^{1/2} \gg a 
\end{equation} 
и 
\begin{equation} \label{eq:10:2.58} 
L_{\tau}=\begin{cases}
a\sqrt{JS/T}\quad & (\text{ФМ}),\\
c/T & (\text{АФМ}).
\end{cases}
\end{equation} 
На масштабе $\xi _{J^{\prime }}$ характер флуктуаций изменяется с двумерных гейзенберговского типа на трехмерные гейзенберговские или двумерные изинговские флуктуации в зависимости от того, какой из параметров доминирует в знаменателе \eqref{eq:10:2.57}~— анизотропия или межплоскостной обмен. С~другой стороны, на масштабе $L_{\tau }$ тип флуктуаций меняется с квантовых на классические. 

Представление производящего функционала \eqref{eq:10:2.56} позволяет произвести учет магнон-магнонного взаимодействия за пределами спин-волновой теории. Возможным способом учета этого взаимодействия, выходящим за рамки низшего порядка теории возмущений, является ренормгрупповой (РГ) анализ. Этот подход ранее успешно применялся для описания классических и квантовых изотропных магнетиков в пространствах размерности $d=2$~\cite{10:047,10:063} и $d=2+\varepsilon $~\cite{10:064,10:065}. В~указанных случаях картина спектра возбуждений слабо отличается от спин-волновой. Так, при $d=2+\varepsilon $ поправки к спектру спиновых волн $\delta E_{\mathbf{q}} \sim |J|\varepsilon \ln q$, температура магнитного перехода $T_{\text{M}} /|J|S^{2} \sim \varepsilon $ и может быть применена стандартная техника $\varepsilon $-разложения. При этом результаты РГ анализа совпадают с результатами $1/N$-разложения в SU($N$)/SU($N-1$) обобщении модели Гейзенберга~\cite{10:059}.

В случае квазидвумерных магнетиков со слабым межплоскостным обменом и~(или) слабой анизотропией типа «легкая ось» температура магнитного перехода также мала в сравнении с $|J|S^{2}$, однако спектр возбуждений может существенно отличаться от спин-волнового. Вне критической области, однако, спиновые флуктуации носят двумерный изотропный характер (по этой причине этот режим далее именуется «\textit{двумерный гейзенберговский режим}»); в этом режиме спектр спиновых возбуждений сохраняет спин-волновой характер и для описания магнитных свойств в этом режиме может быть применен метод РГ. Лишь в узкой критической области вблизи $T_{\text{M}}$ происходит переход от вышеупомянутого двумерного гейзенберговского режима к трехмерному гейзенберговскому (или двумерному изинговскому) \textit{критическому} режиму, в котором картина спиновых волн становится полностью неадекватной. Таким образом, эта область должна рассматриваться с учетом  существенно неспинволновых возбуждений. 

Для применения теоретико-полевых методов производящий функционал \eqref{eq:10:2.56} может быть далее преобразован к виду, удобному для конкретных вычислений; при этом результат определяется температурным режимом, в котором производятся вычисления. В~классическом режиме $T\gg JS$ имеем $L_{\tau } \ll a$ и динамикой поля $\mathbf{n}$ (т.~е. его зависимостью от мнимого времени) можно пренебречь, что приводит к функционалу 
\begin{multline} \label{eq:10:2.59} 
Z_{\text{cl}} [h]=\int D \mathbf{n} \, D\lambda \exp \left\{\frac{\rho _{\text{s}}^{0}}{2T} \sum _{i} \left[\mathbf{n}_{i} \mathbf{n}_{i+\delta _{\Vert }} +\frac{\alpha }{2} \mathbf{n}_{i} \mathbf{n}_{i+\delta _{\bot }} + {} \right. \right.  
\\
\left. \left. {} + \eta n_{i}^{z} n_{i+\delta _{\Vert }}^{z} +\tilde{\zeta }(n_{i}^{z} )^{2} +hn_{i}^{z} + i \lambda (\mathbf{n}_{i}^{2} -1)
\vphantom{\frac{\alpha }{2}}
\right]
\vphantom{\frac{\rho _{\text{s}}^{0}}{2T}}
\right\}
\end{multline} 
с «затравочной» спиновой жесткостью $\rho _{\text{s}}^{0} =|J|S^{2}$. Чтобы получить \eqref{eq:10:2.59} в антиферромагнитном случае, необходимо произвести замену $\mathbf{n}_{i}\to -\mathbf{n}_{i}$, $\lambda _{i}\to -\lambda _{i}$ для одной из двух подрешеток. Таким образом, в классическом случае результаты для~$Z$ идентичны для ферро- и антиферромагнетиков. В~континуальном пределе действие~\eqref{eq:10:2.59} совпадает с действием для классической нелинейной сигма-модели~\cite{10:069}. Однако, если интересоваться термодинамикой в широком интервале температур (не только в критической области), континуальный предел не может быть использован, т.~к. при этом вклад в термодинамические свойства дают не только длинноволновые, но и коротковолновые возбуждения. 

В квантовом случае в силу условия $\xi _{J^{\prime }} \gg a$ можно перейти к континуальному пределу для каждого слоя. Для ферромагнетика удобно использовать представление 
\begin{equation} \label{eq:10:2.60} 
\mathbf{A}(\mathbf{n})=\frac{\mathbf{z}\times \mathbf{n}}{1+(\mathbf{z}\mathbf{n})} 
\end{equation} 
($\mathbf{z}$~— единичный вектор вдоль оси $z$) и ввести двухкомпонентное векторное поле $\pi =\mathbf{n}-(\mathbf{n}\mathbf{z})\mathbf{z}$ описывающее флуктуации параметра порядка. Для квантового антиферромагнетика необходимо использовать процедуру Халдейна~\cite{10:030} (см. также~\cite{10:069}), чтобы проинтегрировать по «быстрым» компонентам поля~$\mathbf{n}$. При этом параметр $\xi _{J^{\prime }} \gg a$ используется для отделения «быстрых» и «медленных» переменных вместо обычно используемой корреляционной длины, равной бесконечности ниже точки перехода. С~помощью указанной процедуры приходим к производящему функционалу квантовой нелинейной сигма-модели 
\begin{multline} \label{eq:10:2.61} 
Z_{\text{AF}} [h]=\int D\sigma  \, D\lambda \exp \left\{-\frac{\rho _{\text{s}}^{0}}{2} \int \limits_{0}^{1/T} d\tau \int d^{2} \mathbf{r} \sum _{i_{z}} \left[\frac{1}{c_{0}^{2}} (\partial _{\tau } \sigma _{i_{z}} )^{2} + {} \right. \right. 
\\
\left. \left. {} +(\nabla \sigma _{i_{z}} )^{2} +\frac{\alpha }{2} (\sigma _{i_{z} +1} -\sigma _{i_{z}} )^{2} -f(\sigma _{i_{z}}^{z} )^{2} +h\sigma _{i_{z}}^{z} +i\lambda (\sigma _{i_{z}}^{2} -1)
\vphantom{\frac{1}{c_{0}^{2}}}
\right]
\vphantom{\int \limits_{0}^{1/T}}
\right\},              
\end{multline} 
где $\sigma _{i_{z}}$~— трехкомпонентное поле единичной длины и $c_{0} =\sqrt{8} JS$~— затравочная скорость спиновых волн. Модель \eqref{eq:10:2.61} обладает O($3$)/O($2$) группой симметрии. В~отличие от случая квантового ферромагнетика, эта модель может быть обобщена на O($N$)/O($N-1$) симметрию с произвольным $N$ путем введения $N$-компонентного векторного поля $\sigma _{i} =\{ \sigma _{1}\ldots \sigma _{N} \}$ и замены $\sigma ^{z}$ на~$\sigma _{N}$.

\subsection{Описание различных температурных режимов в~рамках ренормгруппового подхода и~$1/N$-разложения}
\label{sec:10:2.4}

В двумерном гейзенберговском режиме взаимодействие спиновых волн является существенным, но сами спин-волновые возбуждения являются хорошо определенными. Наличие этого режима является специфической особенностью квазидвумерных систем с малыми значениями межплоскостного обмена и анизотропии. Как можно видеть уже из результатов спин-волновых подходов (разделы~\ref{sec:10:2.1} и~\ref{sec:10:2.2}, в этом режиме имеются логарифмические расходимости в (подрешеточной) намагниченности, определяемые параметрами $\ln (\xi _{J^{\prime }} /L_{\tau } )$ в квантовом и $\ln (\xi _{J^{\prime }} /a)$ в классическом случае. Для суммирования этих расходимостей, представляющих влияние динамического взаимодействия спиновых волн на намагниченность и температуры магнитного перехода, удобно использовать РГ подход~\cite{10:047,10:063,10:064,10:065,10:066}. 

Для применения РГ подхода вводится формальный параметр инфракрасного обрезания $\mu $, так что указанные расходимости заменяются на $\ln [1/(\mu L_{\tau } )]$. Далее рассматриваются температурно-зависящие перенормировочные параметры $\tilde{Z}_{i}$, введенные согласно теоретико-полевой формулировке РГ~\cite{10:064,10:070}
\[
t=t_{\text{R}} Z_{1} ,\quad \pi =\pi _{\text{R}} Z,\quad h=h_{\text{R}} Z_{1} /\sqrt{Z} ,
\] 
\begin{equation} \label{eq:10:2.62} 
f=f_{\text{R}} Z_{2} ,\quad \alpha =\alpha _{\text{R}} Z_{3} ,  
\end{equation} 
являющиеся функциями $\mu $ и опреденными из условия отсутствия  логарифмических расходимостей в перенормированной теории; индекс $R$ соответствует квантово- и температурно-перенормированным величинам. Аналогично классической нелинейной сигма-модели~\cite{10:064}, введение пяти перенормировочных параметров для пяти независимых параметров модели оказывается достаточным для того, чтобы устранить все имеющиеся расходимости (см. также~\cite{10:070}). 

Бесконечно малое изменение $\mu $ генерирует преобразование ренормгруппы. В~двухпетлевом приближении результат для температуры и намагниченности эффективной модели имеет вид
\begin{equation} \label{eq:10:2.63} 
\mu \frac{dt_{\text{r}}}{d\mu } =-(N-2)t_{\text{r}}^{2} -(N-2)t_{\text{r}}^{3} +O(t_{\text{r}}^{4} ),  
\end{equation} 
\begin{equation} \label{eq:10:2.64} 
\mu \frac{d\ln Z}{d\mu } =(N-1)t_{\text{r}} +O(t_{\text{r}}^{3} ).  
\end{equation} 
Перенормировка параметров межплоскостного обмена и анизотропии описывается в однопетлевом приближении (достаточном для рассматриваемого ниже двухпетлевого анализа) функциями 
\begin{equation} \label{eq:10:2.65} 
\mu \frac{d\ln Z_{2}}{d\mu } =-2t_{\text{r}} +O(t_{\text{r}}^{2} ),  
\end{equation} 
\begin{equation} \label{eq:10:2.66} 
\mu \frac{d\ln Z_{3}}{d\mu } =-t_{\text{r}} +O(t_{\text{r}}^{2} ).  
\end{equation} 
Уравнения \eqref{eq:10:2.63}—\eqref{eq:10:2.66} определяют эволюцию параметров модели при РГ преобразовании. 

Для относительной намагниченности $\bar{\sigma }\equiv \bar{S}/S$ получаем уравнение~\cite{10:066}
\begin{equation} \label{eq:10:2.67} 
\bar{\sigma }^{1/\beta _{2}} =1-\frac{t}{2} \left[\ln \frac{2}{u\Delta (f_{t} ,\alpha _{t} )} +2\ln (1/\bar{\sigma }^{1/\beta _{2}} )+2(1-\bar{\sigma }^{1/\beta _{2}} )+O(t/\bar{\sigma }^{1/\beta _{2}} )\right].  
\end{equation} 
где 
\begin{equation} \label{eq:10:2.68} 
\Delta (f,\alpha )=f+\alpha +\sqrt{f^{2} +2\alpha f} ,  
\end{equation} 
$f_{t}$ и $\alpha _{t}$~— температурно-зависящие параметры межплоскостного обмена и анизотропии: 
\begin{equation} \label{eq:10:2.69} 
f_{t} /f_{\text{r}} =\bar{\sigma }_{\text{r}}^{4/(N-1)} \left[1+O(t_{\text{r}} /\bar{\sigma }_{\text{r}}^{1/\beta _{2}} )\right],  
\end{equation} 
\begin{equation} \label{eq:10:2.70} 
\alpha _{t} /\alpha _{\text{r}} =\bar{\sigma }_{\text{r}}^{2/(N-1)} \left[1+O(t_{\text{r}} /\bar{\sigma }_{\text{r}}^{1/\beta _{2}} )\right].  
\end{equation} 
Величина
\begin{equation} \label{eq:10:2.71} 
\beta _{2} =\frac{N-1}{2(N-2)} 
\end{equation} 
есть «критический индекс» подрешеточной намагниченности в рассматриваемом температурном интервале. Она совпадает с пределом критического индекса $\beta _{2+\varepsilon }$ в пространстве размерности $d=2+\varepsilon $~\cite{10:064} при $\varepsilon \to 0$; в физически важном случае $N=3$ имеем $\beta _{2} =1$. Ведущий логарифмический член в квадратных скобках \eqref{eq:10:2.67} соответствует результату ССВТ \eqref{eq:10:2.31}, в то время как другие два члена описывают поправки к этой теории, при этом наиболее важный субведущий логарифмический вклад совпадает с результатами лестничного приближения, рассмотренного в подразделе~\ref{sec:10:2.2}. Результаты \eqref{eq:10:2.69}—\eqref{eq:10:2.71} при $N=3$ совпадают с результатами ССВТ \eqref{eq:10:2.23} и~\eqref{eq:10:2.24}.

Аналогичное рассмотрение может быть произведено для квантовых антиферромагнетиков. В~этом случае подрешеточную намагниченность и температуру фазового перехода удобно выражать через наблюдаемые параметры основного состояния: намагниченность $\bar{S}_{0}$, спиновую жесткость $\rho _{\text{s}}$, скорость спиновых волн $c$, межплоскостной обмен $\alpha _{\text{r}}$ и анизотропию $f_{\text{r}} =(\Delta /\rho _{\text{s}} )\bar{S}_{0}$, где $\Delta $~— щель в энергетическом спектре. Поэтому на первом шаге ренормгруппового преобразования удобно ввести параметры квантовой перенормировки $\tilde{Z}_{i}$ согласно соотношениям
\[
\bar{S}_{0} =\tilde{Z}S,\quad 
g_{0} =g\tilde{Z}_{1} ,\quad 
c_{0} =c\tilde{Z}_{\text{c}} , 
\] 
\[
f=f_{\text{r}} \tilde{Z}_{2} ,\quad 
\alpha =\alpha _{\text{r}} \tilde{Z}_{3} ,
\] 
связывающим наблюдаемые параметры основного состояния $g$, $c$, $\alpha _{\text{r}}$, $f_{\text{r}}$ с (затравочными) параметрами исходной модели $g_{0}$, $c_{0}$, $\alpha $, $f$, где $g=\rho _{\text{s}} /c$ и $g_{0}=\rho _{\text{s}}^{0} /c_{0}$~— безразмерные перенормированная и затравочная константа связи модели~\eqref{eq:10:2.61}. 

В силу неуниверсальности перенормировочных констант $\tilde{Z}_{i}$, т.~е. их зависимости от деталей структуры решетки, они могут быть определены лишь из рассмотрения исходной решеточной (неконтинуальной) версии производящего функционала~\eqref{eq:10:2.56}. Поскольку указанные параметры не содержат логарифмических расходимостей, они могут быть вычислены в спин-волновой теории, являющейся фактически разложением в ряд по $g$ ($g\sim 1/S$ для больших $S$). Для антиферромагнетиков с квадратной решеткой результаты раздела~\ref{sec:10:2.2} приводят к выражениям
\begin{equation} \label{eq:10:2.72} 
\tilde{Z}=1/\tilde{Z}_{1} =\tilde{Z}_{2} =\tilde{Z}_{3}^{1/2} =1-0.197/S, 
\end{equation} 
\[
\tilde{Z}_{\text{c}} =1+0.079/S
\] 
с точностью до членов первого порядка по $1/S$~\cite{10:016,10:039,10:040,10:041,10:042,10:043,10:044,10:045,10:046}. Для учета квантовых перенормировок удобно иметь эквивалент результатов \eqref{eq:10:2.72}, определенный в рамках континуальной модели \eqref{eq:10:2.61}. В~первом порядке по $g$ находим 
\begin{equation} \label{eq:10:2.73} 
\tilde{Z}=1-(N-1)\frac{g\Lambda }{4\pi } +O(g^{2}),
\end{equation} 
\[
\tilde{Z}_{1} =1-(N-2)\frac{g\Lambda }{4\pi } +O(g^{2} ),\quad 
\tilde{Z}_{\text{c}} =1+O(g^{2} ),  
\] 
\[
\tilde{Z}_{2} =1+\frac{g\Lambda }{2\pi } +O(g^{2} ),\quad 
\tilde{Z}_{3} =1+\frac{3g\Lambda }{4\pi } +O(g^{2} ),
\] 
где $\Lambda $~— параметр ультрафиолетового обрезания, необходимый для регуляризации расходимостей, возникающих при вычислении параметров основного состояния. После выполнения квантовой перенормировки \eqref{eq:10:2.73} теория, как мы увидим ниже, становится полностью универсальной, так что термодинамические свойства не зависят от параметра обрезки~$\Lambda $.

Второй шаг РГ преобразования состоит в суммировании логарифмических расходимостей $\ln (\xi _{J^{\prime }} /L_{\tau })$ и аналогичен ферромагнитному случаю. Относительная подрешеточная намагниченность $\bar{\sigma }_{\text{r}} =\bar{\sigma }/\bar{\sigma }_{0}$ квантового квазидвумерного антиферромагнетика может быть определена из  \eqref{eq:10:2.63}—\eqref{eq:10:2.66} в виде~\cite{10:066}:
\begin{equation} \label{eq:10:2.74}
\bar{\sigma }_{\text{r}}^{1/\beta _{2}} =1-\left[\frac{t_{\text{r}}}{2} (N-2)\ln \frac{2}{u_{\text{r}}^{2} \Delta (f_{t} ,\alpha _{t} )} +\frac{2}{\beta _{2}} \ln (1/\bar{\sigma }_{\text{r}} )+2(1-\bar{\sigma }_{\text{r}}^{1/\beta _{2}} )+O(t_{\text{r}} /\bar{\sigma }_{\text{r}}^{1/\beta _{2}} )\right],
\end{equation} 
при этом результаты для температурной перенормировки параметров анизотропии и межплоскостного обмена  имеют тот~же вид \eqref{eq:10:2.69}, \eqref{eq:10:2.70}, что и для ферромагнетика.  

Уравнение для намагниченности в двухпетлевом РГ подходе для классического магнетика может быть получено таким~же образом, как и для квантового случая. Имеем~\cite{10:066}
\begin{equation} \label{eq:10:2.75}
\bar{\sigma }^{1/\beta _{2}}=1-\frac{t_{\text{L}}}{2} \left[ (N-2)\ln \frac{64}{\Delta (f_{t} ,\alpha _{t})} +\frac{2}{\beta _{2}} \ln (1/\bar{\sigma })+2(1-\bar{\sigma }^{1/\beta _{2}} ) +O(t_{\text{L}} /\bar{\sigma }^{1/\beta _{2}} )\right] ,
\end{equation}
где $t_{\text{L}}=tZ_{\text{L}1}^{-1}$, $Z_{\text{L}1}=1-\pi t/2+O(t^{2})$.

Таким образом, РГ подход достаточен для вычисления намагниченности при температурах, не слишком близких к температуре магнитного перехода, при которых спин-волновые возбуждения играют решающую роль, а также позволяет вычислить температуры Кюри (Нееля) с точностью до некоторый постоянной, являющейся универсальной в квантовом случае. Общие скейлинговые результаты для температур Кюри и Нееля могут быть также получены экстраполяцией результатов РГ подхода в критическую область, что для квантовых магнетиков приводит к результатам~\cite{10:028}
\begin{equation} \label{eq:10:2.76} 
t_{\text{C}} =2\left[\ln \frac{2}{u\Delta (f_{\text{c}} ,\alpha _{\text{c}} )} +2\ln (2/t_{\text{C}} )+\Phi _{\text{F}} (\alpha /f)\right]^{-1} ,  
\end{equation} 
\begin{equation} \label{eq:10:2.77} 
t_{\text{N}} =2\left[(N-2)\ln \frac{2}{u_{\text{r}}^{2} \Delta (f_{\text{c}} ,\alpha _{\text{c}} )} +2\ln (2/t_{\text{N}} )+\Phi _{\text{AF}} (\alpha _{\text{r}} /f_{\text{r}} )\right]^{-1} ,  
\end{equation} 
где $\Phi _{\text{F,AF}} (\alpha _{\text{r}} /f_{\text{r}} )\sim 1$~— некоторые функции. Второй член в знаменателе \eqref{eq:10:2.76}, \eqref{eq:10:2.77} представляющий собой поправку к ССВТ, имеет порядок $\ln \ln (2T_{\text{N}}^{2} /\alpha )$ и приводит к существенному понижению температуры Нееля по сравнению с ее значением в ССВТ. Функция $\Phi $ определяется неспинволновыми возбуждениями и не может быть вычислена в рамках РГ подхода. В~квазидвумерном изотропном случае $\Phi $ может быть вычислена с помощью $1/N$-разложения (см.~ниже); более общий случай требует численного анализа (например, квантовым методом Монте-Карло). 

Описание температурной зависимости (подрешеточной) намагниченности в критической области, а также более строгое определение температур магнитного фазового перехода, требует выхода за рамки спин-волновой картины спектра магнитных возбуждений, в частности явного учета вклада продольных спиновых флуктуаций (неявно учитываемых также рассмотренными лестничным приближением и~РГ анализом). 

Учет продольных спиновых флуктуаций, необходимых для правильного описания критической области и полного вычисления температуры Нееля, возможен в рамках $1/N$-разложения~\cite{10:027,10:028}, учитывающего все компоненты спина на равных основаниях и накладывающее условие сохранения спина на узле. Благодаря этому, указанный подход позволяет лучше описать область температур, близкую к температуре магнитного фазового перехода и, в частности, критическую область, хотя он не воспроизводит полностью низкотемпературное поведение намагниченности. Для температуры Нееля $1/N$-разложение воспроизводит РГ результат \eqref{eq:10:2.77}, но при этом позволяет вычислить постоянную $\Phi _{\text{AF}} (0)$, которая не может быть определена в рамках РГ подхода.  

Построение $1/N$-разложения для квантового квазидвумерного антиферромагнетика производится аналогично двумерному случаю~\cite{10:027}, при этом используется обобщение модели Гейзенберга на  модель с O($N$)/O($N-1$) симметрией \eqref{eq:10:2.61}. При~$N=\infty $ указанная модель эквивалентна сферической модели~\cite{10:067}, однако при конечных значениях $N$ она правильно учитывает поправки связанные со спин-спиновым взаимодействием, поскольку не основана на спин-волновой картине спектра. Это обстоятельство приводит к важным преимуществам при температурах, сравнимых с температурой фазового перехода, но ведет к некоторым трудностям при описании низких и промежуточных температур, где возбуждения имеют чисто спин-волновой характер. Таким образом, подходы РГ и $1/N$-разложения в~O($N$)/O($N-1$) модели имеют преимущества в различных температурных областях и взаимно дополняют друг друга.

Рассмотрим вначале изотропный случай. В~связи с наличием дальнего порядка ниже температуры Нееля, произведем сдвиг поля $\sigma =\tilde{\sigma }+\bar{\sigma }$ где $\bar{\sigma }$~— относительная подрешеточная намагниченность $\bar{S}/S$. После интегрирования по $\tilde{\sigma }$ производящий функционал \eqref{eq:10:2.61} принимает вид 
\begin{equation} \label{eq:10:2.78} 
Z[h]=\int D\lambda  \exp (N S_{\text{eff}} [\lambda ,h]) ,
\end{equation} 
\begin{equation} \label{eq:10:2.79} 
S_{\text{eff}} [\lambda ,h]=\frac{1}{2} \ln \det \hat{G}_{0} +\frac{1}{2g} (1-\bar{\sigma }^{2} )\Sp (i\lambda )
+\frac{1}{2g} \Sp \left[(i\lambda \bar{\sigma }-h/\rho _{\text{s}}^{0})\hat{G}_{0} (i\lambda \bar{\sigma }-h/\rho _{\text{s}}^{0} )\right],  
\end{equation} 
где 
\begin{equation} \label{eq:10:2.80} 
\hat{G}_{0} =\left[ \partial _{\tau }^{2} /c_{0}^{2} +\nabla ^{2} +\alpha \Delta _{z} \right] ^{-1} ,  
\end{equation} 
\[
\Delta _{z} \sigma _{i_{z}} (\mathbf{r},\tau )=\sigma _{i_{z} +1} (\mathbf{r},\tau )-\sigma _{i_{z}} (\mathbf{r},\tau ).
\] 
Поскольку $N$ входит в действие \eqref{eq:10:2.79} лишь как множитель в показателе экспоненты, предел $N\to \infty $ соответствует приближению седловой точки функционала $S_{\text{eff}} [\lambda ,h]$, в котором пренебрегается флуктуациями поля~$l$. Этот предел совпадает с так называемой сферической моделью~\cite{10:067}, пренебрегающей связью различных спиновых компонент. При этом физическое условие $\mathbf{S}_{i}^{2} =S(S+1)$ заменяется условием на среднее по узлам значение: 
\begin{equation} \label{eq:10:2.81} 
\sum _{i}\mathbf{S}_{i}^{2} =NS(S+1) .   
\end{equation} 
Такое приближение приводит к резкому упрощению модели, позволяя решить ее точно. Дальнейшие поправки, вычисляемые путем разложения в окрестности седловой точки, дают последовательное улучшение приближения \eqref{eq:10:2.81} по параметру~$1/N$.

При $T<T_{\text{N}}$ седловая точка имеет координаты $i\lambda =0$ и $\bar{\sigma }^{2} \ne 0$. Для определенности далее полагаем, что подрешеточная намагниченность направлена в $N$-ом направлении, т.~е. $\bar{\sigma }^{m} =\bar{\sigma }\delta _{mN}$. Тогда $G^{NN}$ соответствует продольной функции Грина $G_{\text{l}}$, в то время как другие диагональные компоненты~— поперечной функции Грина~$G_{t}$. Условие \eqref{eq:10:2.81} в указанных обозначениях принимает вид 
\begin{equation} \label{eq:10:2.82} 
1-\bar{\sigma }^{2} =\frac{T}{\rho _{\text{s}}^{0}} \sum _{\omega _{n}} \sum _{m} \int \frac{d^{2} \mathbf{k}}{(2\pi )^{2}} \int \frac{dk_{z}}{2\pi } G^{mm} (k,k_{z} ,\omega _{n} ) ,
\end{equation} 
где
$$
G^{mn} (\mathbf{q}, q_{z}, \omega _{n}) = \rho _{\text{s}}^{0} \int d\tau  \, \langle T[\tilde{\sigma }_{\mathbf{q},q_{z}}^{m} (\tau )\tilde{\sigma }_{-\mathbf{q},-q_{z}}^{n} (0)]\rangle .
$$
В пределе $N\to \infty $ имеем для спиновой функции Грина
\begin{equation} \label{eq:10:2.83} 
G_{0} (k,k_{z} ,\omega _{n} )=\left[\omega _{n}^{2} +k^{2} +\alpha (1-\cos k_{z} )\right]^{-1} .  
\end{equation} 
Температура Нееля, определенная из \eqref{eq:10:2.82}, равна 
\begin{equation} \label{eq:10:2.84} 
T_{\text{N}}^{0} =\frac{4\pi \rho _{\text{s}}^{N=\infty }}{N\ln (2T_{\text{N}}^{2} /\alpha c^{2} )} ,  
\end{equation} 
где $\rho _{\text{s}}^{N=\infty } =N(1/g-1/g_{\text{c}})$~— жесткость спиновых волн в нулевом порядке по $1/N$, $g_{\text{c}}=2\pi ^{2} /\Lambda $~— формальный параметр теории. Значение \eqref{eq:10:2.84} в $N/(N-2)$ раз ниже результата ССВТ \eqref{eq:10:2.34} и РГ подхода \eqref{eq:10:2.77}. Это отличие обусловлено недостатком приближения сферической модели, рассматривающей различные спиновые компоненты независимо друг от друга. 

В первом порядке по $1/N$ учитываются наинизшие поправки к условию \eqref{eq:10:2.81}, обусловленные однократным обменом возбуждением поля~$l$, учитывающем связь между различными компонентами спина на узле. Уравнение для намагниченности при $T\gg \alpha ^{1/2}$ и $\ln (2T_{\text{N}}^{2} /\alpha c^{2} )\gg 1$ в первом порядке по $1/N$ имеет вид 
\begin{multline} \label{eq:10:2.85} 
1-\frac{NT}{4\pi \rho _{\text{s}}} \left[ \left(1-\frac{2}{N} \right) \ln \frac{2T^{2}}{\alpha _{\text{r}}} +\frac{3}{N} \ln \frac{4\pi \rho _{\text{s}}}{NTx_{\bar{\sigma }}} -\frac{2}{N} \frac{\ln (2T^{2} /\alpha _{\text{r}} )}{\ln (2T^{2} /\alpha _{\text{r}} )+x_{\bar{\sigma }}} -I_{1} (x_{\bar{\sigma }} )\right] = {} \\
{} = \frac{\bar{\sigma }^{2}}{\bar{\sigma }_{0}^{2}} \left[1+\frac{1}{N} \ln \frac{4\pi \rho _{\text{s}}}{NTx_{\bar{\sigma }}} -I_{2} (x_{\bar{\sigma }} )\right],
\end{multline} 
где $I_{1,2} (x_{\bar{\sigma }} )$~— некоторые функции (см.~\cite{10:028}), $\bar{\sigma }_{0} =\bar{\sigma }(T=0)=\bar{S}_{0} /S$ и $\rho _{\text{s}}$~— подрешеточная намагниченность и спиновая жесткость основного состояния, в нелинейной сигма-модели квантового двумерного антиферромагнетика~\cite{10:027}. Как и в ренормгрупповом подходе, намагниченность подрешетки, выраженная в терминах квантово-перенормированнных величин $\rho _{\text{s}}$, $\bar{\sigma }_{0}$ и $\alpha $, не зависит от параметра обрезания $\Lambda $, т.~е. является универсальной величиной. 

В области температур 
\begin{equation} \label{eq:10:2.86} 
NT/4\pi \rho _{\text{s}} \ll \bar{\sigma }^{2} /\bar{\sigma }_{0}^{2} ,  
\end{equation} 
не слишком близких к точке магнитного перехода $x_{\bar{\sigma }} \gg 1$, так что функции $I_{1} (x_{\bar{\sigma }} )$ и $I_{2} (x_{\bar{\sigma }} )$ порядка $1/x_{\bar{\sigma }}$, т.~е. малы по сравнению с остальными вкладами. Уравнение для намагниченности при температурах, определяемых условием \eqref{eq:10:2.86}, имеет вид 
\begin{multline} \label{eq:10:2.87} 
(\bar{\sigma }/\bar{\sigma }_{0} )^{1/\beta _{2}} [1-I_{2} (x_{\bar{\sigma }} )] = {} \\
{} =1-\frac{NT}{4\pi \rho _{\text{s}}} \left[\left(1-\frac{2}{N} \right) \ln \frac{2T^{2}}{\alpha _{\text{r}}} +\frac{3}{N} \ln \frac{\bar{\sigma }_{0}^{2}}{\bar{\sigma }^{2}} - \frac{2}{N} \frac{\ln (2T^{2} /\alpha _{\text{r}} )}{\ln (2T^{2} /\alpha _{\text{r}} )+x_{\bar{\sigma }}} -I_{1} (x_{\bar{\sigma }} )\right],  
\end{multline} 
где 
\begin{equation} \label{eq:10:2.88} 
x_{\bar{\sigma }} =\frac{4\pi \rho _{\text{s}}}{(N-2)T} \frac{\bar{\sigma }^{2}}{\bar{\sigma }_{0}^{2}} .  
\end{equation} 
Результат \eqref{eq:10:2.87} имеет вид, сходный с результатом ренормгруппы \eqref{eq:10:2.74}, в то~же время отличаясь от него коэффициентом при субведущем члене $\ln (\bar{\sigma }_{0} /\bar{\sigma })$ ($6/N$ вместо $3/b_{2}$). Это отличие связано с тем, что $1/N$-разложение не способно вполне корректно описать двумерный гейзенберговский режим. Указанное различие проявляется, однако, лишь в членах в намагниченности порядка $1/N^{2}$, лежащих за пределами точности первого порядка по $1/N$ и не приводит к существенному отклонению результатов $1/N$-разложения от РГ результатов в области низких и промежуточных температур (см. обсуждение экспериментальных результатов в разделе~\ref{sec:10:2.5}). В~то~же время, давая качественно правильное описание двумерного режима, уравнение \eqref{eq:10:2.87} позволяет правильно описать и подрешеточную намагниченность в режиме, переходном к критическому.

Для исследования критического режима рассмотрим температуры, близкие к~$T_{\text{N}}$, так что $\bar{\sigma }$ достаточно мало, чтобы удовлетворить условию 
\begin{equation} \label{eq:10:2.89} 
\bar{\sigma }^{2} /\bar{\sigma }_{0}^{2} \ll (N-2)T/4\pi \rho _{\text{s}} ,  
\end{equation} 
при котором $x_{\bar{\sigma }} \ll 1$. После разложения \eqref{eq:10:2.87} вблизи $T=T_{\text{N}}$, $x_{\bar{\sigma }}=0$, имеем 
\begin{equation} \label{eq:10:2.90} 
1-\frac{T}{T_{\text{N}}} =\frac{\bar{\sigma }^{2}}{\bar{\sigma }_{0}^{2}} \left[1+\frac{1}{N} \ln \frac{4\pi \rho _{\text{s}}}{(N-2)T_{\text{N}}} +\frac{8}{\pi ^{2} N} \ln x_{\bar{\sigma }} -A_{0} \right],  
\end{equation} 
где $A_{0} =2.8906/N$. Преобразуя логарифмические члены в степени, находим результат для подрешеточной намагниченности в критической области,
\begin{equation} \label{eq:10:2.91} 
\frac{\bar{\sigma }}{\bar{\sigma }_{0}} =\left[\frac{4\pi \rho _{\text{s}}}{(N-2)T_{\text{N}}} \right]^{(\beta _{3} /\beta _{2} -1)/2} \left[\frac{1}{1-A_{0}} \left(1-\frac{T}{T_{\text{N}}} \right)\right]^{\beta _{3}} , 
\end{equation}
где 
\begin{equation} \label{eq:10:2.92} 
\beta _{3} =\frac{1}{2} \left(1-\frac{8}{\pi ^{2} N} \right) 
\end{equation} 
— критический индекс намагниченности. При $N=3$ имеем $\beta _{3} \simeq 0.36$, что совпадает с результатом $1/N$-разложения в $\phi ^{4}$-модели~\cite{10:071} при $d=3$, в согласии с гипотезой универсальности. Уравнение для температуры Нееля $T_{\text{N}}$ имеет вид 
\begin{equation} \label{eq:10:2.93} 
T_{\text{N}} =4\pi \rho _{\text{s}} \left[(N-2)\ln \frac{2T_{\text{N}}^{2}}{\alpha _{\text{r}}} +3\ln \frac{4\pi \rho _{\text{s}}}{(N-2)T_{\text{N}}} -0.0660\right]^{-1} . 
\end{equation} 

Исследование спектра магнитных возбуждений в точке магнитного фазового перехода, определяющегося собственно-энергетической частью $\Sigma (k,k_{z} ,0)$ при $T=T_{\text{N}}$, позволяет определить температурно-перенормированный параметр межплоскостной анизотропии
\begin{equation} \label{eq:10:2.94} 
\alpha _{\text{c}} =\alpha _{\text{r}} \left(1+\frac{1.0686}{N} \right)\left[\frac{(N-2)T_{\text{N}}}{4\pi \rho _{\text{s}}} \right]^{1/(N-2)} . 
\end{equation} 
Так~же как в ССВТ (см. раздел~\ref{sec:10:2.2} перенормированное значение параметра межплоскостного обмена в $T_{\text{N}}$ ниже, чем его низкотемпературное значение, но конкретное выражение при $N=3$ отличается от результата ССВТ численным множителем, примерно равным~$1.3$. Критический индекс для асимптотики корреляционной функции в точке перехода определяется выражением
\begin{equation} \label{eq:10:2.95} 
\eta =8/(3\pi ^{2} N) 
\end{equation} 
При $N=3$ имеем $\eta \simeq 0.09$. Значения остальных индексов могут быть определены из \eqref{eq:10:2.92} и \eqref{eq:10:2.95} с помощью скэйлинговых соотношений.

\subsection{Теоретическое описание экспериментальных данных намагниченности и~температур Нееля слоистых систем}
\label{sec:10:2.5}

Суммируем результаты в практически важном случае $N=3$. В~спин-волновой и двумерной областях, т.~е. при 
\begin{equation} \label{eq:10:2.96} 
\bar{\sigma }_{\text{r}} \gg \frac{T}{4\pi \rho _{\text{s}}} ,\quad \Gamma \gg \Delta  
\end{equation} 
результат РГ для относительной (подрешеточной) намагниченности имеет вид
\begin{equation} \label{eq:10:2.97} 
\bar{\sigma }_{\text{r}} =1-\frac{T}{4\pi \rho _{\text{s}}} \left[\ln \frac{2\Gamma (T)}{\Delta (f_{t} ,\alpha _{t} )} +2\ln (1/\bar{\sigma }_{\text{r}} )+2(1-\bar{\sigma }_{\text{r}} )\right],  
\end{equation} 
где функция $\Delta (f,\alpha )$ определена в \eqref{eq:10:2.68}, температурно-перенормированные значения межплоскостного обмена и параметра анизотропии 
\begin{equation} \label{eq:10:2.98} 
\frac{f_{t}}{f_{\text{r}}} =\left( \frac{\alpha _{t}}{\alpha _{\text{r}}} \right)^{2} =\bar{\sigma }_{\text{r}}^{2} 
\end{equation} 
и величины $\Gamma (T),\bar{\sigma }_{\text{r}} ,f_{\text{r}} ,\alpha _{\text{r}} ,\rho _{\text{s}}$ определены в таблице~\ref{tab:10:01}.  Уравнение для $T_{\text{M}}$ имеет вид 
\begin{equation} \label{eq:10:2.99} 
T_{\text{M}} =4\pi \rho _{\text{s}} \left[\ln \frac{2\Gamma (T_{\text{M}} )}{\Delta (f_{\text{c}} ,\alpha _{\text{c}} )} +2\ln \frac{4\pi \rho _{\text{s}}}{T_{\text{M}}} +\Phi (f/\alpha )\right]^{-1} ,  
\end{equation} 
где $\Phi (x)$~— некоторая (универсальная в квантовом случае) функция порядка единицы, $f_{\text{c}}$ и $\alpha _{\text{c}}$~— параметры межплоскостного обмена и анизотропии при $T=T_{\text{M}}$, причем
\begin{equation} \label{eq:10:2.100} 
\frac{f_{\text{c}}}{f_{\text{r}}} =\left( \frac{\alpha _{\text{c}}}{\alpha _{\text{r}}} \right)^{2} =\left( \frac{T_{\text{M}}}{4\pi \rho _{\text{s}}} \right)^{2} .  
\end{equation} 
Так как $T_{\text{M}} /4\pi \rho _{\text{s}} \sim 1/\ln (1/\Delta )\ll 1$, температурные перенормировки важны для правильного описания экспериментальных данных. В~частности, параметры межплоскостного обмена и анизотропии, измеренные при различных температурах, могут значительно отличаться.

\begin{table}[bp]
\caption{Параметры уравнений для подрешеточной намагниченности \eqref{eq:10:2.97} и температуры магнитного перехода \eqref{eq:10:2.99} для различных случаев, $Z_{\text{L}1} =Z_{\text{L}2} =Z_{\text{L}3} =1-T/8\pi \rho _{\text{s}}^{0}$}
\label{tab:10:01}
\vspace*{.5em}%
\newcolumntype{Y}{>{\centering\arraybackslash}X}%
\begin{tabularx}{\textwidth}{|l|Y|Y|Y|Y|Y|}
\hline 
                     & $\Gamma (T)$   & $\bar{\sigma }_{\text{r}}$   & $\rho _{\text{s}}$            & $f_{\text{r}}$                  & $\alpha _{\text{r}}$          \\ \hline 
квантовый АФМ        & $T^{2} /c^{2}$ & $\bar{S}/\bar{S}_{0}$ & $\gamma S\bar{S}_{0}$  & $f\bar{S}_{0}^{2}/S^{2}$ & $\alpha \bar{S}_{0}/S$ \\ \hline 
квантовый ФМ         & $T/JS$         & $\bar{S}/S$           & $\rho _{\text{s}}^{0}$        & $f$                      & $\alpha $              \\ \hline 
классический ФМ, АФМ & $32$           & $\bar{S}/S$           & $\rho _{\text{s}}^{0} Z_{\text{L}1}$ & $fZ_{\text{L}2}^{-1}$           & $\alpha Z_{\text{L}3}^{-1}$   \\ \hline 
\end{tabularx}
\normalsize%
\end{table}

При $\alpha =0$ имеем 
\begin{equation} \label{eq:10:2.101} 
\bar{\sigma }_{\text{r}} =1-\frac{T}{4\pi \rho _{\text{s}}} \left[\ln \frac{\Gamma (T)}{f_{\text{r}}} +4\ln (1/\bar{\sigma }_{\text{r}} )+2(1-\bar{\sigma }_{\text{r}} )\right],  
\end{equation} 
\begin{equation} \label{eq:10:2.102} 
T_{\text{M}} =4\pi \rho _{\text{s}} \left[\ln \frac{\Gamma (T_{\text{M}} )}{f_{\text{r}}} +4\ln \frac{4\pi \rho _{\text{s}}}{T_{\text{M}}} +\Phi (0)\right].  
\end{equation} 
При $f=0$ 
\begin{equation} \label{eq:10:2.103} 
\bar{\sigma }_{\text{r}} =1-\frac{T}{4\pi \rho _{\text{s}}} \left[\ln \frac{2\Gamma (T)}{\alpha _{\text{r}}} +3\ln (1/\bar{\sigma }_{\text{r}} )+2(1-\bar{\sigma }_{\text{r}} )\right],  
\end{equation} 
\begin{equation} \label{eq:10:2.104} 
T_{\text{M}} =4\pi \rho _{\text{s}} \left[\ln \frac{2\Gamma (T_{\text{M}} )}{\alpha _{\text{r}}} +3\ln \frac{4\pi \rho _{\text{s}}}{T_{\text{M}}} +\Phi (\infty )\right].  
\end{equation} 
Результаты ССВТ \eqref{eq:10:2.31}—\eqref{eq:10:2.39} отличаются от соответствующих ренормгрупповых результатов \eqref{eq:10:2.101}—\eqref{eq:10:2.104} заменой $4(3)\to 2(1)$ для коэффициента во втором члене в квадратных скобках соответственно в анизотропном двумерном (изотропном квазидвумерном) случаях. Таким образом, роль поправок к ССВТ более важна в изотропном квазидвумерном магнетике, чем в двумерном анизотропном. 

Результат $1/N$-разложения в O($N$) модели вне критической области, точнее при 
\begin{equation} \label{eq:10:2.105} 
\bar{\sigma }_{\text{r}}^{2} > \frac{T}{4\pi \rho _{\text{s}}} ,\quad \Gamma \gg \Delta  
\end{equation} 
в первом порядке по $1/N$ имеет вид 
\begin{equation} \label{eq:10:2.106} 
[1-I_{2} (x_{\bar{\sigma }} )] \bar{\sigma }_{\text{r}} = 1-\frac{T}{4\pi \rho _{\text{s}}} \left[\ln \frac{2\Gamma (T)}{\Delta (f_{\text{r}} ,\alpha _{\text{r}} )} 
+2B_{2} \ln (1/\bar{\sigma }_{\text{r}} )+2(1-\bar{\sigma }_{\text{r}}^{2} )+I_{1} (x_{\bar{\sigma }} )\right],                       
\end{equation} 
где $x_{\bar{\sigma }} =4\pi \rho _{\text{s}} \bar{\sigma }_{\text{r}}^{2} /T$, $B_{2}=3+f_{\text{r}} /\sqrt{f_{\text{r}}^{2} +2\alpha _{\text{r}} f_{\text{r}}}$, $\Delta $ определено в \eqref{eq:10:2.68}, $I_{1,2} (x)$~— некоторые функции с асимптотикой $1/x$ при больших $x$, другие величины приведены в таблице~\ref{tab:10:01}. В~частных случаях $\alpha =0$ и $f=0$ коэффициент при втором члене в квадратных скобках в \eqref{eq:10:2.106} вдвое больше чем для РГ результатов \eqref{eq:10:2.101}, \eqref{eq:10:2.103}. В~спин-волновой и двумерной температурных областях это различие является недостатком приближения первого порядка по~$1/N$. В~то~же время $1/N$-разложение обеспечивает более правильное описание температурной области, переходной к критическому поведению, и критической области. Из-за различия условий на переходную область \eqref{eq:10:2.96} и \eqref{eq:10:2.105} уравнения для $T_{\text{M}}$ имеют вид, одинаковый в обоих подходах. В~изотропном случае ($f=0$) результат $1/N$-разложения для подрешеточной намагниченности в критической области
\begin{equation} \label{eq:10:2.107} 
\bar{\sigma }_{\text{r}} =\left(\frac{4\pi \rho _{\text{s}}}{T_{\text{N}}} \right)^{(\beta _{3} -1)/2} \left[\frac{1}{1-A_{0}} \left(1-\frac{T}{T_{\text{N}}} \right)\right]^{\beta _{3}} ,  
\end{equation} 
где $A_{0} =0.9635$ и $\beta _{3} \simeq 0.36$.

\begin{figure}[bp]
\centering
\includegraphics{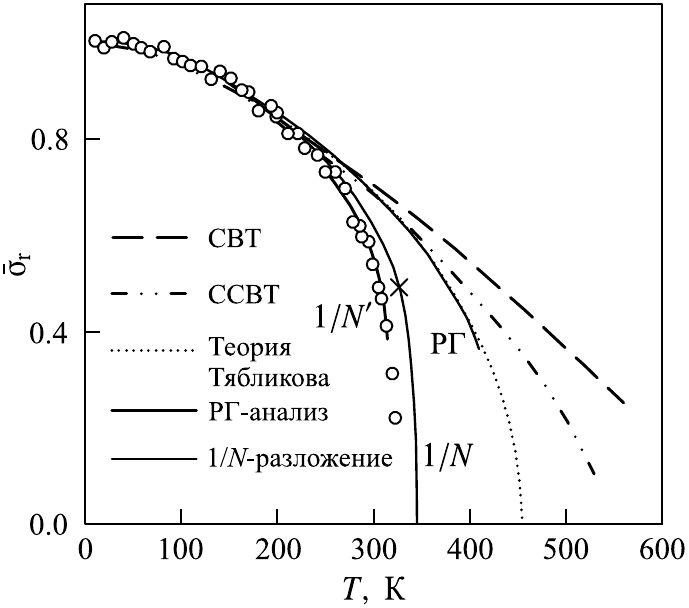}
\caption{Теоретические температурные зависимости относительной намагниченности подрешетки $\bar{\sigma }_{\text{r}}$ в различных приближениях: спин-волновых теориях, РГ подходе \eqref{eq:10:2.97} и $1/N$-разложении в O($N$) модели (уравнения \eqref{eq:10:2.106} и \eqref{eq:10:2.107}) и экспериментальные точки для La$_{2}$CuO$_{4}$~\cite{10:002}. Кривая РГ приведена вплоть до температуры, где производная $\partial \bar{\sigma }_{\text{r}} /\partial T$ расходится. Кривая, обозначенная как $1/N^{\prime }$, ближе к экспериментальным данным в переходной температурной области благодаря включению анизотропии, определенной из условия равенства $T_{\text{M}}$ его экспериментальному значению (см. обсуждение в тексте)}
\label{fig:10:005}
\end{figure}

Рассмотрим теперь применение полученных результатов для анализа экспериментальных данных. Одним из хорошо изученных слоистых соединений является La$_{2}$CuO$_{4}$~\cite{10:007,10:082}. Значение перенормированного параметра обмена для этого соединения, $\gamma |J|\simeq 1850$~К может быть определено из экспериментальных данных для спин-волнового спектра при низких температурах~\cite{10:083}, в то время как значение межплоскостного обмена $\alpha _{\text{r}} =1\cdot 10^{-3}$ может быть найдено из сравнения намагниченности в ССВТ с экспериментальной зависимостью~$\bar{\sigma }_{\text{r}} (T)$ при низких температурах~\cite{10:028,10:066}. На рисунке~\ref{fig:10:005} представлены экспериментальная температурная зависимость намагниченности подрешетки в La$_{2}$CuO$_{4}$~\cite{10:082}, результаты спин-волновых приближений (СВТ, ССВТ и теории Тябликова) для этого соединения, РГ подхода и $1/N$-разложения. Результат для температуры Нееля $1/N$-разложения первого порядка $T_{\text{N}} =345$~К, что значительно ниже всех спин-волновых приближений и находится в хорошем согласии с экспериментальным значением $T_{\text{N}}^{\text{exp}} =325$~К.

РГ подход правильно описывает зависимость~$\bar{\sigma }_{\text{r}} (T)$ в спин-волновой области $(T<300K)$ и области двумерных флуктуаций (которая очень узка при вышеприведенном малом значении~$\alpha $), в то время как при более высоких температурах этот подход переоценивает $\bar{\sigma }$. С~другой стороны, кривая $1/N$-разложения расположена ближе всего к экспериментальным данным и правильно описывает критическое поведение. Результаты численного решения уравнения~\eqref{eq:10:2.106} в температурной области~\eqref{eq:10:2.105} и зависимости~\eqref{eq:10:2.107} в критической области, совпадают в точке $T=330$~К, отмеченной крестиком. Различие между теоретической и экспериментальной кривыми в температурной области $320$~К$<T<340$~К может быть обусловлено влиянием анизотропии. При фиксированном $\Delta $ и $B_{2}$, определенном из наилучшего совпадения с экспериментальными данными при промежуточных температурах (см. рис.~\ref{fig:10:005}), находим значения $\alpha _{\text{r}} =1\cdot 10^{-4}$, $f_{\text{r}} =5\cdot 10^{-4}$. Таким образом, рассматриваемый подход дает возможность оценить относительную роль межплоскостного обмена и магнитной анизотропии в слоистых соединениях. Отметим, что альтернативное объяснение различия между теоретическим и экспериментальным результатами, основанное на рассмотрении циклического 4-х спинового взаимодействия, было предложено в работе~\cite{10:084}.

\begin{figure}[htbp]
\centering
\includegraphics{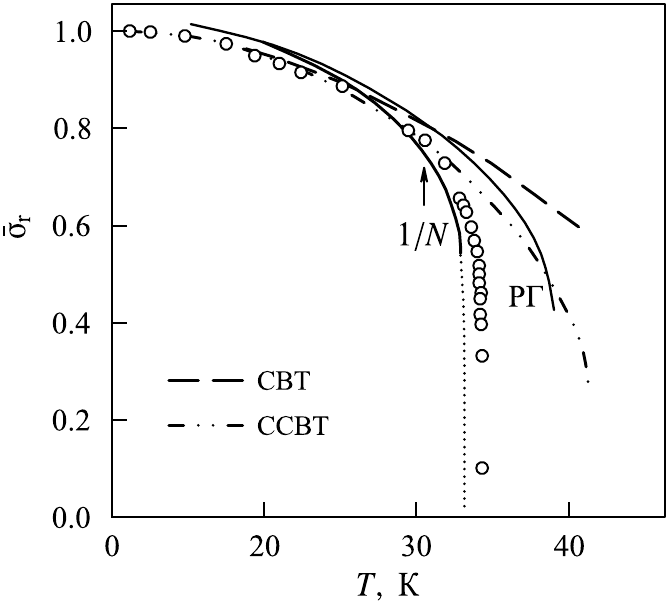}
\caption{Относительная намагниченность подрешетки $\bar{\sigma }_{\text{r}} (T)$ для K$_{2}$NiF$_{4}$  (точки) по сравнению со стандартной спин-волновой теорией (пунктир), ССВТ (штрих-пунктир), РГ подходом и результатом решения уравнения \eqref{eq:10:2.106} в промежуточной температурной области (сплошная линия). Короткий пунктир показывает экстраполяцию результата $1/N$-разложения на изинговскую критическую область. Граница между областью с флуктуациями двумерного типа и поведения переходного к критическому отмечена стрелкой}
\label{fig:10:006}
\end{figure}

В слоистых перовскитах K$_{2}$NiF$_{4}$, Rb$_{2}$NiF$_{4}$ и K$_{2}$MnF$_{4}$ магнитная анизотропия, как известно, является более важной, чем межплоскостной обмен. Соединение K$_{2}$NiF$_{4}$ имеет спин $S=1$, из данных нейтронного рассеяния следует $|J|=102$~К и $T_{\text{N}}^{\text{exp}} =97.1$~К~\cite{10:003}. На рисунке~\ref{fig:10:006} показана экспериментальная зависимость~$\bar{\sigma }(T)$~\cite{10:001} и результаты спин-волновых подходов, РГ подхода и $1/N$-разложения. Значение $f_{\text{r}} =0.0088$ было получено из сравнения результата намагниченности ССВТ с экспериментальными данными при низких температурах (это значение хорошо согласуется с экспериментальным $f_{\text{r}} =0.0084$~\cite{10:003}). В~спин-волновом и двумерном флуктуационном температурных интервалах  ($T<80$~К) кривые, соответствующие $1/N$-разложению и РГ подходу, располагаются несколько выше, чем экспериментальные точки, поскольку $T^{2} /f_{\text{r}} c^{2}$ в этой области не велико. В~то~же время кривая $1/N$-разложения находится в хорошем численном согласии с экспериментальными данными. Процедура экстраполяции к изинговскому критическому поведению дает $T_{\text{N}} =91.4$~К, причем ширина критической изинговской области составляет $1$~К. Отметим, что учет членов порядка $1/x_{\bar{\sigma }}$ в \eqref{eq:10:2.106} приводит к значению $T_{\text{N}} =92.7$~К. В~переходной к критическому поведению области $80<T<90$~К теоретическая O($3$)~кривая для K$_{2}$NiF$_{4}$, в отличие от случая La$_{2}$CuO$_{4}$ лежит слегка ниже экспериментальной. Этот факт может быть приписан влиянию межплоскостного обмена. Определение соответствующих параметров в переходной области приводит к значениям $\alpha _{\text{r}} =0.0017$, $f_{\text{r}} =0.0069$, которые соответствуют $T_{\text{N}} =97$~К и затравочным параметрам $\alpha |J|=0.1$~К, $\zeta |J|=0.76$~К. Соответствующие экспериментальные данные для $\alpha $ отсутствуют, поэтому сравнение с экспериментом в данном случае затруднительно.

\begin{figure}[htbp]
\centering
\includegraphics{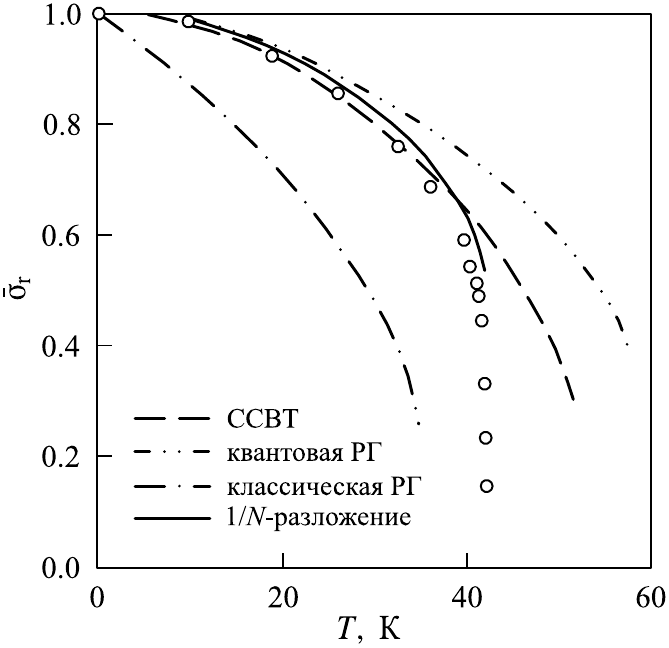}
\caption{Экспериментальная зависимость $\bar{\sigma }_{\text{r}} (T)$ для K$_{2}$MnF$_{4}$ (точки) по сравнению с результатами ССВТ (пунктирная линия), квантовым РГ анализом (две точки-пунктир), классическим РГ анализом (штрих-пунктир) и решением \eqref{eq:10:2.106} (сплошная линия)}
\label{fig:10:007}
\end{figure}

\begin{figure}[htbp]
\centering
\includegraphics{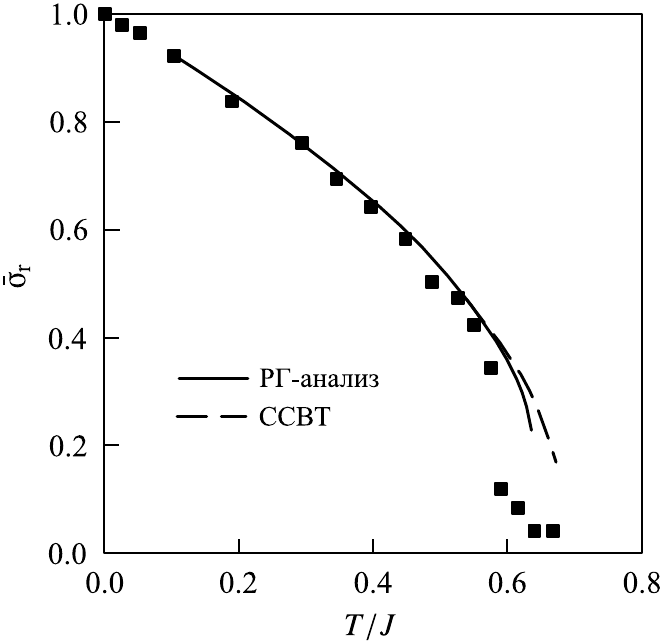}
\caption{Результаты ренормгруппового подхода (сплошная линия) и ССВТ (пунктирная линия) для относительной намагниченности $\bar{\sigma }$ классического анизотропного двумерного магнетика ($\zeta =0$, $\eta =0.001$) в сравнении с результатами вычисления методом Монте-Карло~\cite{10:085}. РГ и ССВТ кривые показаны до температуры, где $\partial \bar{\sigma }/\partial T=\infty $}
\label{fig:10:008}
\end{figure}

Соединение Rb$_{2}$NiF$_{4}$ обладает сильной магнитной анизотропией: согласно~\cite{10:003} $|J|=82$~К, $|J|f_{\text{r}} =3.45$~К, $T_{\text{N}}^{\text{exp}} =94.5$~К. Сравнение экспериментальной зависимости $\bar{\sigma }_{\text{r}} (T)$ с результатами ССВТ при низких температурах приводит к значению параметра анизотропии $f_{\text{r}} =0.046$, в хорошем согласии с вышеприведенным экспериментальным значением. Из \eqref{eq:10:2.102} следует $T_{\text{N}} =95.5$~К, что также находится близко к экспериментальным данным для температуры Нееля.

\begin{table}[htbp]
\caption{Экспериментальные параметры и температуры магнитного перехода слоистых магнетиков и~соответствующие теоретические значения $T_{\text{M}}$ в стандартной спин-волновой теории (СВТ), самосогласованной спин-волновой теории (ССВТ) и $1/N$-разложении}
\label{tab:10:02}
\vspace*{.5em}%
\newcolumntype{Y}{>{\centering\arraybackslash}X}%
\begin{tabularx}{\textwidth}{|l|Y|Y|Y|Y|Y|c|Y|Y|}
\hline 
Соединение        & $S$   & $J$, К  & $J^{\prime }$, К & $\eta $  & $T_{\text{СВТ}}$, К & $T_{\text{ССВТ}}$, К & $T_{1/N}$, К & $T_{\text{эксп}}$, К \\ \hline 
La$_{2}$CuO$_{4}$ & $1/2$ & $1600$  & $0.8$            & $0$      & $672$               & $537$                & $343$        & $325$                \\ \hline 
K$_{2}$NiF$_{4}$  & $1$   & $102$   & $0$              & $0.0088$ & $160$               & $125$                & $90.0$       & $97.1$               \\ \hline 
Rb$_{2}$NiF$_{4}$ & $1$   & $82$    & $0$              & $0.046$  & $180$               & $118$                & $88.4$       & $94.5$               \\ \hline 
K$_{2}$MnF$_{4}$  & $5/2$ & $8.4$   & $0$              & $0.015$  & $74.8$              & $52.1$               & $42.7$       & $42.1$               \\ \hline 
CrBr$_{3}$        & $3/2$ & $12.38$ & $1$              & $0.024$  & $79.2$              & $51.2$               & $39.0$       & $40.0$               \\ \hline 
\end{tabularx}
\normalsize%
\end{table}

Соединение K$_{2}$MnF$_{4}$ имеет спин $S=5/2$ и поэтому представляет собой промежуточную ситуацию  между квантовым и классическим случаями. Параметры обмена и анизотропии $|J|=8.4$~К, $|J|f_{\text{r}} =0.13$~К могут быть найдены из данных нейтронного рассеяния~\cite{10:003}. Рисунок~\ref{fig:10:007} показывает сравнение результатов различных подходов с экспериментальными данными для этого соединения. Можно видеть, что $1/N$-разложение приводит к результатам, хорошо описывающим экспериментальную ситуацию во всем интервале температур. В~то~же время экспериментальные точки расположены между квантовой и классической РГ кривыми, причем квантовое приближение является более удовлетворительным. Это подтверждает квантовый характер поправок к намагниченности даже при относительно большой величине спина. В~рассматриваемом случае ССВТ, правильно учитывающая возбуждения на масштабе постоянной решетки, приводит к лучшим результатам по сравнению с РГ подходом. Таким образом, аккуратное рассмотрение систем с большим спином в рамках континуальных моделей требует численного расчета интегралов по импульсам и суммирования по мацубаровским частотам. 

Рисунок~\ref{fig:10:008} показывает сравнение результатов ССВТ и РГ подхода для намагниченности классического магнетика с вычислениями методом Монте-Карло~\cite{10:085}. Можно видеть, что за исключением узкой критической области, кривая РГ довольно точна, хотя при этом пренебрегается топологическими возбуждениями. Область применимости РГ подхода в классическом случае более широка, чем в квантовом случае, так что нет необходимости использовать $1/N$-разложение для описания переходной и критической области. 

Описанные результаты сравнения теоретических и экспериментальных данных по слоистым перовскитам суммированы в таблице~\ref{tab:10:02} и показывают, что РГ подход и $1/N$-разложение в O($N$) модели приводят к количественно правильным результатам температур магнитного перехода и намагниченности этих систем, находящихся в хорошем согласии с экспериментальными данными.

\section{Квазидвумерные магнетики с~анизотропией типа «легкая плоскость»}
\label{sec:10:3}

Другой важный класс низкоразмерных систем~— двумерные системы с анизотропией типа «легкая плоскость», обсуждавшиеся во Введении. Классическая двумерная $XY$~модель, соответствующая предельному случаю сильной легкоплоскостной анизотропии, была подробно изучена в ранних работах~\cite{10:086,10:087,10:088}. В~указанных работах было продемонстрировано, что элементарными возбуждениями в этой модели являются топологические вихревые структуры и существует переход Березинского—Костерлица—Таулеса, связанный с диссоциацией вихревых пар при температуре
\begin{equation} \label{eq:10:3.1} 
T_{\text{BKT}} =\frac{\pi }{2} |J|S^{2} .  
\end{equation} 
При этой~же температуре степенная зависимость корреляционной функции спинов от расстояния изменяется на экспоненциальную (в квантовой $XY$~модели ситуация более сложная, поскольку должны быть учтены не только поперечные, но и продольные компоненты спина). 

Более физически реальная ситуация, однако, описывается двумерной моделью Гейзенберга \eqref{eq:10:2.1} со слабой анизотропией типа «легкая плоскость», т.~е. $\eta ,\zeta <0$ и $|\eta |, |\zeta |, \alpha \ll 1$ (для удобства в дальнейшем сделаем замену $\eta \to -\eta $, $\zeta \to -\zeta $). В~этом случае спин-волновые возбуждения при низких температурах играют определяющую роль в температурной зависимости (подрешеточной) намагниченности. Как и в случае «легкая ось», при температурах, не слишком низких по сравнению с температурой магнитного фазового перехода, необходим правильный учет динамического взаимодействия спиновых волн. 

При слабой анизотропии «легкая плоскость», однако, переход Березинского—Костерлица—Таулеса предшествует магнитному фазовому переходу. При этом благодаря существованию «квазидальнего» порядка при $T<T_{\text{BKT}}$ включение сколь угодно слабого межплоскостного обмена приводит к появлению магнитного перехода выше $T_{\text{BKT}}$. Простое выражение для температуры Березинского—Костерлица—Таулеса, полученное в пределе малой анизотропии, имеет вид~\cite{10:089}
\begin{equation} \label{eq:10:3.2} 
T_{\text{BKT}} =\frac{4\pi |J|S^{2}}{\ln (\pi ^{2} /\eta )} .  
\end{equation} 
Как и для изотропных и легкоосных магнетиков, формула \eqref{eq:10:3.2} недостаточна для количественного описания экспериментальных данных. 

Аналогично магнетикам с анизотропией типа «легкая ось», можно ожидать, что термодинамические свойства этих систем, за исключением узкой окрестности $T_{\text{BKT}}$, определяются возбуждениями спин-волнового типа и для учета влияния взаимодействия спиновых волн при температурах вне критической области вновь может быть применен метод ренормгруппы~\cite{10:090}.

РГ анализ снова выполняется на основе функционала \eqref{eq:10:2.56}. В~классическом случае (т.~е. в пренебрежении динамической частью действия, содержащей производную по времени), имеется два типа возбуждений: поле~$n_{y}$ описывает бесщелевые возбуждения в плоскости, а поле~$n_{z}$~— возбуждения с поворотом спина поперек плоскости, обладающие щелью в энергетическом спектре. Разложение~\eqref{eq:10:2.56} по~$n_{y,z}$ (ось квантования (подрешеточной) намагниченности предполагается вдоль~$x$) приводит в ведущем порядке по~$1/S$ к действию 
\begin{equation} \label{eq:10:3.3} 
L_{\text{st}}=\frac{1}{2} S^{2} \int \limits_{0}^{1/T} d\tau \sum _{\mathbf{k}} [(J_{0} -J_{\mathbf{k}} )\pi _{y\mathbf{k}} \pi _{y,-\mathbf{k}} +(J_{0} -J_{\mathbf{k}} -\eta J_{\mathbf{k}} )\pi _{z\mathbf{k}+\mathbf{Q}} \pi _{z,-\mathbf{k}-\mathbf{Q}} ],  
\end{equation} 
где $\mathbf{Q}$~— волновой вектор магнитной структуры и вектор $\mathbf{n}$ был представлен в виде $n=\{ \sigma ,\pi _{y} ,\pi _{z} \}$. Температура Костерлица—Таулеса может быть вычислена методом, аналогичным вычислению температуры магнитного перехода в случае анизотропии типа «легкая ось». Уравнения ренормгруппы в двумерном гейзенберговском режиме остаются теми же, что и для анизотропии «легкая ось» с заменой $\eta \to -\eta $, $\zeta \to -\zeta $. Вычисление, температуры Костерлица—Таулеса приводит к результату~\cite{10:090}
\begin{equation} \label{eq:10:3.4} 
t_{\text{BKT}} =\left[\ln (\mu _{0} /\sqrt{\eta })+2\ln (2/t_{\text{BKT}} )+C\right]^{-1},
\end{equation} 
где $C$~— универсальная постоянная.

\begin{figure}[htbp]
\centering
\includegraphics{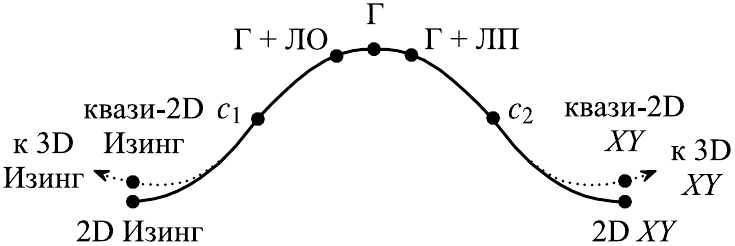}
\caption{Схематическая картина РГ траекторий в слоистых магнетиках. Левая сторона: преобразование от двумерной модели Гейзенберга с анизотропией «легкая ось» (Г\,+\,ЛО) к двумерной модели Изинга. Правая сторона: преобразование от двумерной модели Гейзенберга с анизотропией «легкая плоскость» (Г\,+\,ЛП) к двумерной $XY$~модели. Точки перегиба $c_{1}$, $c_{2}$ отмечают переходные области. Пунктирные линии~— для соответствующих квазидвумерных моделей}
\label{fig:10:009}
\end{figure}

Для вычисления корреляционной длины при температурах, больших температуры Костерлица—Таулеса и определения температуры магнитного фазового перехода в присутствии межплоскостного обмена, произведем «сшивку» результатов, полученных в O($3$) режиме с известными результатами для эффективной классической $XY$~модели. Действительно, даже если исходная модель~— квантовая, на масштабах $\mu \ll \sqrt{\eta } \ll L_{\tau }^{-1}$ эффективная $XY$~модель является классической, поскольку $L_{\tau }$ определяет характерный масштаб, отделяющий квантовые флуктуации от классических (см. раздел~\ref{sec:10:2.3} и рис.~\ref{fig:10:009}). Таким образом, все квантовые эффекты уже учтены на масштабах $\mu \gg \sqrt{\eta }$, где поведение РГ траекторий~— гейзенберговское.

Стандартная система РГ уравнений~\cite{10:087,10:088} двумерной классической $XY$~модели может быть записана в используемых обозначениях в виде 
\[
\mu \frac{d(1/t_{\mu } )}{d\mu } =32\pi ^{2} y_{\mu }^{2} ,
\] 
\begin{equation} \label{eq:10:3.5} 
\mu \frac{dy_{\mu }}{d\mu } =-y_{\mu } \left(2-\frac{1}{2t_{\mu }} \right) .
\end{equation} 
Необходимо отметить, что константой связи для системы вихрей является не $t$ (как для спиновых волн), а $y=\exp (-E_{0} /T)$ где $E_{0}$~— энергия вихря. Пусть $\mu _{1} \ll \sqrt{\eta }$~— масштаб, где осуществляется переход от уравнений \eqref{eq:10:2.63}—\eqref{eq:10:2.66} к уравнениям \eqref{eq:10:3.5}. Обозначим соответствующую эффективную температуру $t_{1} \equiv t_{\mu _{1}}$, константу связи вихрей $y_{1} \equiv y_{\mu _{1}}$ так что
\begin{equation} \label{eq:10:3.6} 
\frac{1}{t_{1}} =\frac{1}{t} -\ln \frac{\mu _{0}}{\sqrt{\eta }} +2\ln \frac{t}{t_{1}} +\Phi (\mu _{1} ) , 
\end{equation} 
\[
y_{1} =\frac{1}{4\pi } \left[\frac{\mu }{2} \frac{d\Phi (\mu )}{d\mu } \right]_{\mu =\mu _{1}}^{1/2} . 
\] 
Тогда решение уравнений \eqref{eq:10:3.5} для $t\geqslant t_{\text{BKT}}$ имеет вид  
\begin{equation} \label{eq:10:3.7} 
\frac{1}{t_{\mu }} =4+2C_{1} \tan \left(C_{1} \ln \frac{\mu }{\mu _{1}} +C_{2} \right) , 
\end{equation} 
где 
\[
C_{1} =\frac{\sqrt{(8\pi y_{1} t_{1} )^{2} -(4t_{1} -1)^{2}}}{2t_{1}},
\] 
\begin{equation} \label{eq:10:3.8} 
\tan C_{2} =\frac{1-4t_{1}}{\sqrt{(8\pi y_{1} t_{1} )^{2} -(4t_{1} -1)^{2}}} . 
\end{equation} 
Температура Костерлица—Таулеса $T_{\text{BKT}}$ определяется уравнением сепаратрисы 
\begin{equation} \label{eq:10:3.9} 
8\pi y_{1} =\frac{1}{t_{1}} -4,\quad t=t_{\text{BKT}} 
\end{equation} 
отделяющей низко- и высокотемпературные фазы. Для достаточно малых $\mu $ имеем $\Phi (\mu )\to \const $, $d\Phi (\mu )/d\mu \to 0$, и для $t_{\text{BKT}} =T_{\text{BKT}} /(2\pi JS^{2} )$ (или $T_{\text{BKT}} /(2\pi \rho _{\text{s}} )$ в АФМ-случае) воспроизводим результат \eqref{eq:10:3.4} с $C=4-6\ln 2-\Phi (\mu \to 0)$. 

В критической области выше $t_{\text{BKT}}$,
\begin{equation} \label{eq:10:3.10} 
\frac{t_{\text{BKT}}^{-1} -t^{-1}}{8\pi } \ll 1, 
\end{equation} 
выражение для корреляционной длины, полученное из \eqref{eq:10:3.7}, имеет вид 
\begin{equation} \label{eq:10:3.11} 
\xi =\frac{1}{\mu _{1}} \exp \left(-\frac{C_{2}}{C_{1}}\right) \simeq \frac{1}{\sqrt{\eta }} \exp \left(\frac{A}{2\sqrt{t_{\text{BKT}}^{-1} -t^{-1}}} \right) 
\end{equation} 
сходный с результатом для классической $XY$~модели ($A$~— некоторая константа). При условии, обратном \eqref{eq:10:3.10}, имеем стандартное гейзенберговское поведение~\cite{10:047} 
\begin{equation} \label{eq:10:3.12} 
\xi = \frac{C_{\xi }}{\mu _{0}}t\exp \frac{1}{t}. 
\end{equation} 

В присутствии межплоскостного обмена, при достаточно низких температурах возникает магнитный порядок. Из-за топологических эффектов, температура перехода при малом межплоскостном обмене стремится к $T_{\text{BKT}}$, а не к нулю. В~случае $\alpha \ll \eta $ выберем $\mu _{1}$ таким, что $\alpha ^{1/2} \ll \mu _{1} \ll \eta ^{1/2}$. В~терминах РГ преобразования, при $\mu =\mu _{1}$ необходимо рассматривать квазидвумерную эффективную $XY$~модель с постоянной решетки $\mu _{0} /\mu _{1}$ и межплоскостным обменом $(\mu _{0} /\mu _{1} )^{2} \alpha _{1}$, где 
\begin{equation} \label{eq:10:3.13} 
\alpha _{1} \equiv \alpha _{\mu _{1}} =\alpha \frac{t}{t_{1}} . 
\end{equation} 
При РГ преобразовании эта модель преобразуется к трехмерной $XY$~модели. Описание этой части РГ преобразования затруднительно  вследствие сложной геометрии вихревых петель в трехмерном пространстве. Вместо прямого вычисления РГ траекторий, используем те~же самые аргументы как в разделе~\ref{sec:10:2.4} для квазидвумерного случая с анизотропией «легкая ось». Температура перехода может быть определена из требования, чтобы корреляционная длина модели без межплоскостного обмена ($\alpha =0$) совпадала с характерным масштабом перехода от двумерной к трехмерной $XY$~модели, $1/\alpha _{1}^{1/2}$ (в единицах постоянной решетки). Тогда находим для критической температуры $t_{\text{c}} =T_{\text{C}} /(2\pi JS^{2} )$ (или $T_{\text{N}} /(2\pi \rho _{\text{s}} )$) в случае $\alpha \ll \eta $
\begin{equation} \label{eq:10:3.14} 
t_{\text{c}} =\left[\ln \frac{\mu _{0}}{\sqrt{\eta }} +2\ln \frac{2}{t_{\text{BKT}}} +C-\frac{A^{2}}{\ln ^{2} (\eta /\alpha )} \right]^{-1} .
\end{equation} 
Последний член в знаменателе определяет разницу между $t_{\text{c}}$ и $t_{\text{BKT}}$. Так как этот член может быть не слишком мал, по нему не производится разложение результата~\eqref{eq:10:3.14}. 

Результат \eqref{eq:10:3.14} качественно правилен вплоть до $\alpha $ порядка $\eta $ (в этом случае, последний член в знаменателе приводит только к перенормировке константы $C$). В~обратном случае $\alpha \gg \eta $ поправки к результату РГ для квазидвумерных магнетиков вследствие анизотропии типа «легкая плоскость» определяются как 
\begin{equation} \label{eq:10:3.15} 
t_{\text{c}} =\left[\ln \frac{\mu _{0}}{\sqrt{\alpha }} +2\ln \frac{2}{t_{\text{c}}} +C^{\prime } +O\left(\frac{\eta ^{1/\psi }}{\alpha ^{1/\psi }} \right)\right]^{-1} , 
\end{equation} 
где $\psi =\nu _{3} (2-\gamma _{\eta })$~— критический индекс перехода (кроссовера) между изотропным и анизотропным поведением, $\nu _{3}$~— соответствующий критический индекс трехмерной модели Гейзенберга, и $\gamma _{\eta }$~— аномальная размерность параметра анизотропии трехмерной модели Гейзенберга. Результат $\varepsilon $-разложения в анизотропной $\phi ^{4}$-модели в размерности $4-\varepsilon $ при $\varepsilon =1$ есть $\psi \simeq 0.83$~\cite{10:070}. Для антиферромагнетика, согласно \eqref{eq:10:2.93} постоянная $C^{\prime }\simeq -0.066$. В~отличие от \eqref{eq:10:3.14}, последний член в знаменателе \eqref{eq:10:3.15} имеет степенную зависимость от параметра анизотропии. Это есть следствие факта, что корреляционная длина в трехмерной модели Гейзенберга имеет степенное поведение с температурой (значение $\nu _{3}$ конечно). По этой причине поправка в знаменателе \eqref{eq:10:3.15} мала и для малой анизотропии ей можно пренебречь. 

Обратимся теперь к экспериментальной ситуации. Наиболее экспериментально исследованная система с анизотропией типа «легкая плоскость»~— соединение K$_{2}$CuF$_{4}$ является ферромагнетиком со спином $S=1/2$, $T_{\text{BKT}} =5.5$~К, $T_{\text{C}} =6.25$~К и параметрами $J=20$~К, $\eta =0.04$, $\alpha =6\cdot 10^{-4}$~\cite{10:003}. При подстановке этих значений в \eqref{eq:10:3.4} и \eqref{eq:10:3.14} можно определить $C\simeq -0.5$ и $A\simeq 3.5$. Эти значения констант могут быть проверены на других системах.

Другой пример квазидвумерного ФМ с анизотропией «легкая плоскость»~— соединение NiCl$_{2}$ с $S=1$. Согласно~\cite{10:003} его параметры~— $J=20$~К, $\eta =8\cdot 10^{-3}$ и $\alpha =5\cdot 10^{-5}$. Используя значения $A$ и $C$, определенные для K$_{2}$CuF$_{4}$, находим $T_{\text{BKT}} =17.4$~К и $T_{\text{C}} =18.7$~К в согласии с экспериментальными данными (оба значения $T_{\text{BKT}}$ и $T_{\text{C}}$ лежат в области $18-20$~К). В~то~же время вычисления с ведущей логарифмической точностью согласно \eqref{eq:10:3.2} приводят к $T_{\text{BKT}} =35.3$~К, что вдвое больше экспериментального значения. 

Соединение BaNi$_{2}$(PO$_{4}$)$_{2}$ согласно~\cite{10:003} является  антиферромагнетиком с $S=1$, $|J|=22.0$~К и анизотропией $\eta =0.05$, $\alpha =1\cdot 10^{-4}$. Вычисление дает~\cite{10:099} $T_{\text{BKT}} =23.0$~К, что совпадает с экспериментальным значением и $T_{\text{N}}=24.3$~К, снова в хорошем согласии с $T_{\text{N}}^{\text{exp}} =24.5\pm 1$~К. Несмотря на то, что для этого соединения $T_{\text{BKT}} \sim |J|S$, этот случай также должен рассматриваться как квантовый в соответствии с критерием квантового режима $(T/JS)^{2} \ll 32$ (см. раздел~\ref{sec:10:2.2}).

\section{Слоистые изотропные антиферромагнетики с~треугольной решеткой}
\label{sec:10:4}

Особый случай слоистых систем представляют собой квазидвумерные магнетики с треугольными слоями, в которых существенную роль играют фрустрации. Примерами таких систем являются антиферромагнетики VCl$_{2}$~\cite{10:072} и VBr$_{2}$ с крайне малым обменом между слоями.

Основное состояние в модели Гейзенберга для треугольной решетки в приближении ближайших соседей~— неколлинеарное антиферромагнитное, причем намагниченность подрешетки заметно подавлена квантовыми флуктуациями. При включении обмена между вторыми соседями может происходить фазовый переход в состояние спиновой жидкости. Нелинейная сигма-модель неколлинеарного антиферромагнетика~\cite{10:07I}
\begin{multline} \label{eq:10:4.1} 
S_{\text{n}} =\int \limits_{0}^{1/T} d\tau  \int d^{2} x \left[\frac{1}{2} \left(\chi _{\text{out}}^{0} (|\partial _{\tau } \mathbf{e}_{1}|^{2} +|\partial _{\tau } \mathbf{e}_{2}|^{2} )-[2\chi _{\text{out}}^{0} -\chi _{\text{in}}^{0} ](\mathbf{e}_{1} \partial _{\tau } \mathbf{e}_{2})^{2} \right) + {} \right. \\
\left. {} + \frac{1}{2} \left(\rho _{\text{out}}^{0} (|\nabla \mathbf{e}_{1}|^{2} +|\nabla \mathbf{e}_{2}|^{2} )-[2\rho _{\text{out}}^{0} -\rho _{\text{in}}^{0} ](\mathbf{e}_{1} \nabla \mathbf{e}_{2})^{2} \right)\right],
\end{multline} 
характеризуется двумя спиновыми жесткостями $\rho _{\text{in},\text{out}}^{0}$ и восприимчивостями $\chi _{\text{in},\text{out}}^{0}$ для спиновых возбуждений в плоскости упорядочения и с выходом из указанной  плоскости. Первому типу возбуждений соответствует голдстоуновская (бесщелевая) точка спектра, соответствующая повороту всех спинов в плоскости на один и тот~же угол, а второму~— две голдстоуновских точки, соответствующие вращению плоскости в целом в двух возможных направлениях. 

\begin{figure}[htbp]
\centering
\includegraphics{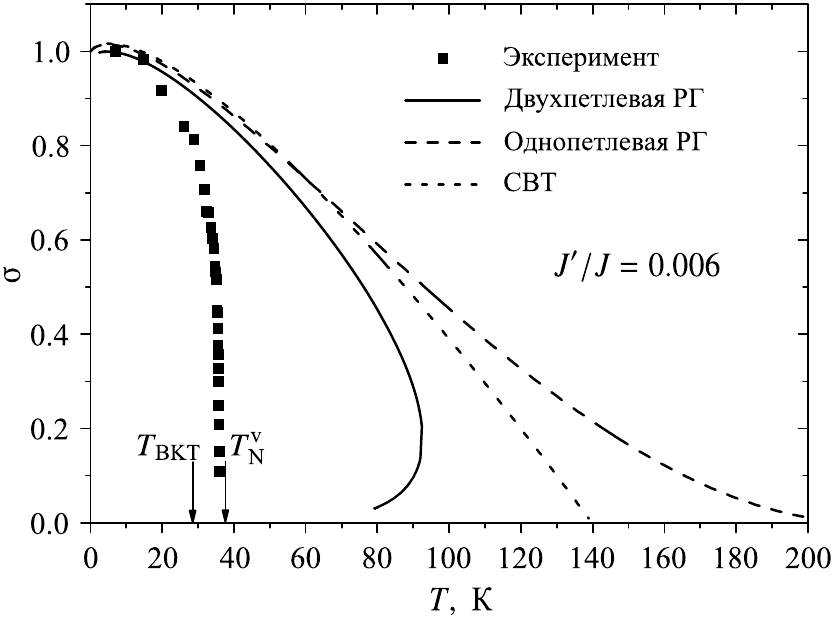}
\caption{Теоретические температурные зависимости относительной намагниченности подрешетки $\sigma $  для слоистой треугольной структуры  с $J^{\prime}/J=0.006$ в рамках спин-волновых и ренормгрупповых подходов. Квадратиками обозначены результаты по нейтронному рассеянию для VCl$_{2}$~\cite{10:072}. $T_{\text{BKT}}$~— температура активации вихрей, $T_{\text{N}}^{\text{v}}$~— оценка точки Нееля с учетом вихрей}
\label{fig:10:010}
\end{figure}

Ситуация в квазидвумерном случае также является специфичной. Ренормгрупповые уравнения могут быть записаны в виде, аналогичном коллинеарному случаю~\cite{10:073}:
\begin{multline*}
\Lambda \frac{db}{d\Lambda } =\frac{(1+b)^{2}}{2y} \left[(N-1) b+3-N-{} \vphantom{\frac{1+b}{2y}} \right.\\
\left. {}-\frac{1+b}{2y} [7(N-3)+b((N-1)b+10-4N))]\right]+O(y^{-3} ) , 
\end{multline*}
\[
\Lambda \frac{dy}{d\Lambda } =(N-2)(1+b)^{2} \left[\frac{1}{2} +\frac{1}{8} \frac{(1+b)^{2}}{y} \right]+O(y^{-2} ),
\] 
\[
\Lambda \frac{d\ln \alpha _{\text{out}}}{d\Lambda } =\frac{3+b(2+b)}{2y} +O(y^{-2} ),\quad 
\Lambda \frac{d\ln \alpha _{\text{in}}}{d\Lambda } =\frac{N-(N-2)b^{2}}{2y} +O(y^{-2} ),
\]
\begin{equation} \label{eq:10:4.2} 
\Lambda \frac{d\ln \bar{\sigma }}{d\Lambda } =-\frac{1}{2} \frac{(N-2)(1+b)+1}{y} +O(y^{-3} ) ,
\end{equation} 
где $y=\rho _{\text{in}}/T$, $b=\rho _{\text{in}}/\rho _{\text{out}}-1$, $\alpha _{\text{in},\text{out}}$~— параметры межплокостного обмена, соответствующие двум спиновым жесткостям, $N$~— число спиновых компонент ($N=3$ в физическом случае). Однако из рисунка~\ref{fig:10:010} видно, что как спин-волновая теория, так и ренормгрупповое рассмотрение~\cite{10:073} являются недостаточными, поскольку не приводят к удовлетворительному согласию с экспериментальными данными. Причина этого расхождения состоит в наличии топологических $Z_{2}$-вихрей в нелинейной сигма-модели \eqref{eq:10:4.1}. Являясь нетривиальными топологическими конфигурациями, они не могут быть учтены в ренормгруппе, которая учитывает только локальные свойства спиновых конфигураций на решетке.

Топологические вихри в изотропной модели Гейзенберга на треугольной решетке до некоторой степени похожи на вихри в $XY$~модели~\cite{10:074,10:075}. Однако в нашем случае спиновые волны не являются свободными, и поэтому их нельзя точно исключить, проинтегрировав по соответствующим степеням свободы.

\begin{table}[ht]
\caption{Параметры и температуры Нееля слоистых антиферромагнетиков с треугольной решеткой}
\label{tab:10:00}
\vspace*{.5em}%
\newcolumntype{Y}{>{\centering\arraybackslash}X}%
\begin{tabularx}{\textwidth}{|l|Y|Y|Y|Y|Y|}
\hline 
            & $J$, К & $J^{\prime}/J$ & $\Delta T/T_{\text{N}}$ & $T_{\text{N}}^{\text{exp}}$, К & $T_{\text{N}}^{\text{v}}$, К \\ \hline 
VBr$_{2}$   & $32$ & $0.06$   & $0.6$ & $29$ & $56.6$ \\ \hline 
VCl$_{2}$   & $44$ & $0.006$  & $0.3$ & $36$ & $39.8$ \\ \hline 
LiCrO$_{2}$ & $80$ & $0.0013$ & $0.2$ & $62$ & $62.5$ \\ \hline 
\end{tabularx}
\normalsize%
\end{table}

Учет вихрей может быть выполнен с помощью метода Монте-Карло~\cite{10:076,10:077,10:078}. Такие вычисления (как и теоретические предсказания~\cite{10:074}) показывают, что в двумерной модели на достаточно большом расстоянии вихри связаны друг с другом логарифмическим кулоновским взаимодействием, так что при $T>T_{\text{BKT}}$ корреляционная длина $\xi (T)$ имеет вид Костерлица—Таулесса~\cite{10:075}
\begin{equation} \label{eq:10:4.3} 
\xi (T)=A \exp \left[b/\sqrt{T-T_{\text{BKT}}} \right] , 
\end{equation} 
где $T_{\text{BKT}}=0.28 JS^{2}$ и $b=0.77$~\cite{10:077}. Температура Нееля в квазидвумерном случае может быть определена как температура кроссовера между 2D и квази-2D режимами, $\xi (T_{\text{N}} )\approx a\sqrt{J/J^{\prime }}$. Таким образом, находим температуру Нееля с учетом вихрей~\cite{10:073}
\begin{equation} \label{eq:10:4.4} 
T_{\text{N}}^{\text{v}} \approx T_{\text{BKT}} +2.37 JS^{2} \ln ^{-2} \left(\frac{2J^{\prime }}{J} \right).
\end{equation} 
В таблице~\ref{tab:10:00} приведены экспериментальные и вычисленные по этой формуле значения точки Неля для слоистых антиферромагнетиков с треугольной решеткой ($S=3/2$, $\Delta T=T_{\text{N}}^{\text{v}}-T_{\text{BKT}}$).
Видно, что в случае малого межслоевого обмена согласие является очень хорошим.

\section{Квазиодномерные изотропные антиферромагнетики}
\label{sec:10:5}

\subsection{Модель и~самосогласованный спин-волновой подход}
\label{sec:10:5.1}

Для описания квазиодномерных систем может быть также использована модель Гейзенберга \eqref{eq:10:2.1}. Ниже рассматривается простейший случай изотропных антиферромагнетиков ($\eta =\zeta =0$) со спином $S=1/2$ и малым межцепочечным обменом $|J^{\prime }|\ll J$. При этом гамильтониан удобно записать в виде 
\begin{equation} \label{eq:10:5.1} 
\mathscr{H}=J\sum _{n,i} \mathbf{S}_{n,i} \mathbf{S}_{n+1,i} +\frac{1}{2} J^{\prime } \sum _{n,\langle ij\rangle } \mathbf{S}_{n,i} \mathbf{S}_{n,j} ,  
\end{equation} 
где $n$ нумерует узлы в цепочке, $i$, $j$~— индексы цепочек, $J>0$ и $J^{\prime }$~— внутри- и межцепочечный обменные интегралы соответственно.

\begin{figure}[bp]
\centering
\includegraphics{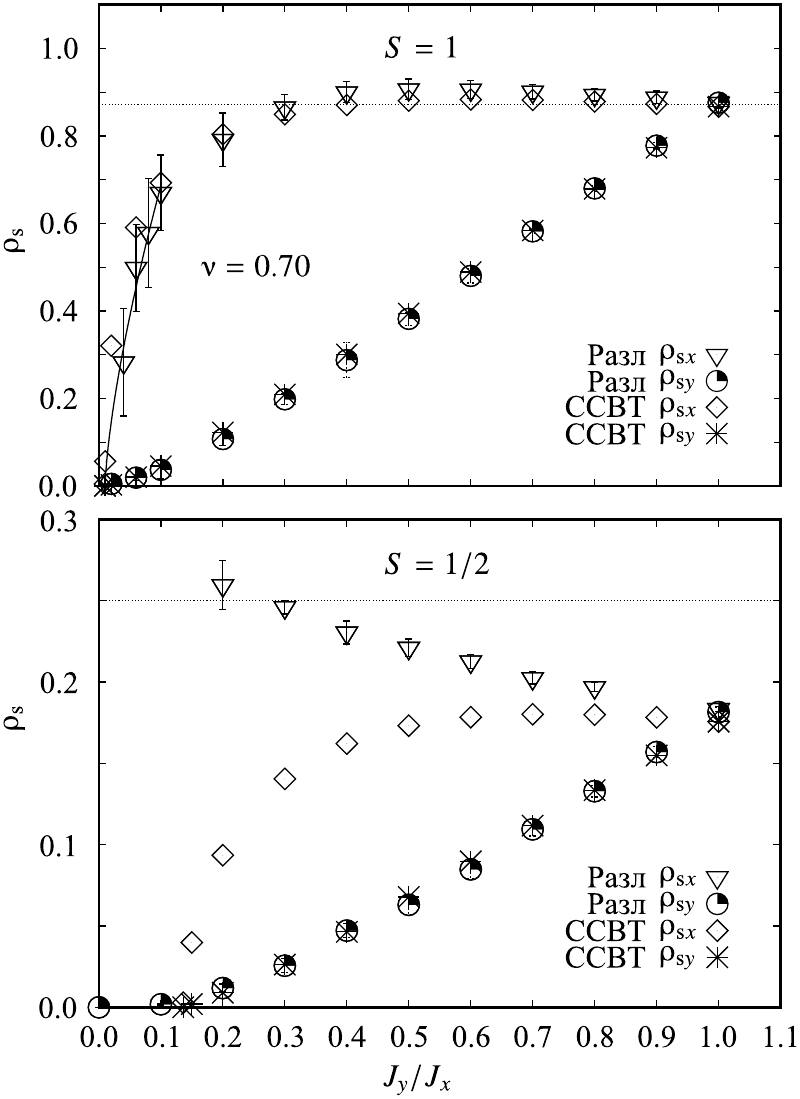}
\caption{Результаты численных разложений (Разл) и самосогласованной спин-волновой теории (ССВТ) для спиновой жесткости вдоль~$x$ и~$y$ направлений квазиодномерной модели Гейзенберга с $S=1$ (верхний рисунок) и $S=1/2$ (нижний рисунок) как функция $J^{\prime}/J$. Тонкая линия показывает значение спиновой жесткости в одномерной модели}
\label{fig:10:011}
\end{figure}

Чтобы применить самосогласованный спин-волновой подход, разделим решетку на подрешетки $A$ и~$B$ и используем представление Дайсона—Малеева \eqref{eq:10:2.3} для операторов спина в каждой подрешетке. Расцепление четырех-бозонных членов приводит к результату       
\begin{equation} \label{eq:10:5.2} 
\mathscr{H}_{\text{SSWT}} =\sum _{i,\delta } \gamma _{\delta } (B_{i}^{\dagger } B_{i} -B_{i+\delta }^{\dagger } B_{i} ), 
\end{equation} 
где 
\[
B_{i}=\begin{cases}
a_{i}, & i\in A,\\
b_{i}^{\dagger },\quad & Ji\in B,
\end{cases}
\] 
$\delta =x, y$ соответствуют ближайшим соседям в направлении $x$ и $y$, 
\begin{equation} \label{eq:10:5.3} 
\gamma _{\delta } =J_{\delta } (\bar{S}+\langle a_{i} b_{i+\delta } \rangle ) 
\end{equation} 
параметры ближнего порядка, $J_{x}=J$, $J_{y}=J^{\prime }$ и $\bar{S}$~— подрешеточная намагниченность. Диагонализуя гамильтониан \eqref{eq:10:5.2}, находим уравнения ССВТ при $T=0$
\[
\gamma _{\delta } =\bar{S}+\sum _{k} \frac{\Gamma _{k}}{2E_{k}} \cos k_{\delta } ,
\] 
\begin{equation} \label{eq:10:5.4} 
\bar{S}=S+\frac12-\sum _{k} \frac{\Gamma _{0}}{2E_{k}} , 
\end{equation} 
где спектр спиновых волн имеет вид
\begin{equation} \label{eq:10:5.5} 
E_{k} =\sqrt{\Gamma _{0}^{2} -\Gamma _{k}^{2}} ,
\end{equation} 
с 
\begin{equation} \label{eq:10:5.6} 
\Gamma _{k} = 2(\gamma _{x} \cos k_{x} +\gamma _{y} \cos k_{y}) 
\end{equation} 
и $\Gamma _{0} \equiv \Gamma _{k=0}$ (мы предполагаем, что основное состояние~— антиферромагнитно упорядочено). Аналогично рассмотрению раздела~\ref{sec:10:2.1} параметры $\gamma _{\delta }$ просто связаны с функцией корреляции вращения на само-соседних участках  $\gamma _{\delta } /J_{\delta } =|\langle S_{i} S_{i+\delta } \rangle |^{1/2}$. Константа спиновой жесткости вдоль направлений $x$ и~$y$ выражается через эти параметры как
\begin{equation} \label{eq:10:5.7} 
\rho _{\text{s}\delta } =S\bar{S}\gamma _{\delta } . 
\end{equation} 
Результаты вычислений согласно уравнениям \eqref{eq:10:5.4}—\eqref{eq:10:5.7} показаны на рисунке~\ref{fig:10:011}, где также произведено их сравнение с результатами численного анализа модели \eqref{eq:10:5.1}, проведенного на основе построения разложения по параметру анизотропии типа «легкая ось» с последующим переходом к изотропному пределу~\cite{10:079}. 

Для спина $S=1$ спин-волновые результаты близки к численным во всем диапазоне $0<J^{\prime }/J<1$. Дальний антиферромагнитный порядок исчезает при значении анизотропии $J^{\prime }/J=0.01$ в согласии с результатами других подходов (см. например~\cite{10:080}). Спиновая жесткость исчезает в точке магнитного квантового фазового перехода с критическим индексом $n=0.7$ в согласии с результатами масштабного анализа~\cite{10:081}; cпин-волновая теория предсказывает несколько большее значение критического индекса~$n=1$. 

В то~же время, при $S=1/2$ два рассмотренных метода дают качественно различные результаты для спиновой жесткости вдоль цепочек $\rho _{\text{s}x}$: в то время как спиновая жесткость в спин-волновой теории уменьшается с уменьшением $J^{\prime }/J$ (аналогично случаю $S=1$), численный анализ приводит к увеличению $\rho _{\text{s}x}$. При этом количественное несоответствие между теорией спиновых волн и численным анализом видно уже при маленьких анизотропиях, с увеличением анизотропии различие становится качественным. Расхождение между результатами спин-волновой теории и численными данными демонстрирует недостаточность спин-волновой теории для описания квазиодномерных систем со спином и возникает в связи с наличием топологических возбуждений, не учитываемым в рамках этой теории.

\subsection{Процедура бозонизации} 
\label{sec:10:5.3}

Для учета топологических возбуждений в случае спина $S=1/2$ необходим выход за рамки спин-волновой теории. Наиболее удобной процедурой, позволяющей это сделать, является процедура «бозонизации». С~этой целью, спиновые операторы представляются как функции ферми-операторов, которые затем выражаются существенно нелинейным образом через бозонные операторы (см., например,~\cite{10:091}). Полученные соотношения между  спиновыми и бозевскими операторами содержат информацию как о спин-волновых, так и топологических возбуждениях, и имеют вид 
\begin{equation} \label{eq:10:5.8} 
\mathbf{S}_{n,i} =\mathbf{J}_{i} (x)+(-1)^{n} \mathbf{n}_{i} (x),  
\end{equation} 
где 
\[
J_{i}^{z} (x)=\frac{\beta }{2\pi } \partial _{x} \varphi _{i} (x),
\] 
\begin{equation} \label{eq:10:5.9} 
J_{i}^{\pm } (x)=\frac{\lambda }{\pi } \exp [\pm i\beta \theta _{i} (x)] \cos \beta \varphi _{i} (x),  
\end{equation} 
и 
\[
n_{i}^{z} (x)=\frac{\lambda }{\pi } \cos \beta \varphi _{i} (x),
\] 
\begin{equation} \label{eq:10:5.10} 
n_{i}^{\pm } (x)=\frac{\lambda }{\pi } \exp [\pm i\beta \theta _{i} (x)],  
\end{equation} 
$\lambda $~— постоянная масштаба обратной постоянной решетки, $\beta =\sqrt{2\pi }$.

Гамильтониан \eqref{eq:10:5.1}, записанный в терминах бозе-операторов $\varphi _{i} (x)$ имеет вид  
\begin{multline} \label{eq:10:5.11} 
\mathscr{H}=\frac{v}{2} \sum _{i} \int dx \left[\Pi _{i}^{2} +(\partial _{x} \varphi _{i} )^{2} \right]+g_{\text{u}} \sum _{i} \int dx \cos 2\beta \varphi _{i} - {}
\\
{}-\frac{J^{\prime } \lambda ^{2}}{2\pi ^{2}} \sum _{i,\delta _{\bot }} \int dx [\cos (\beta \varphi _{i} )\cos (\beta \varphi _{i+\delta _{\bot }} ) +\cos \beta (\theta _{i+\delta _{\bot }} -\theta _{i} )],  
\end{multline} 
где $v=\pi J/2$, $\Pi _{i}$ является импульсом, канонически сопряженным с $\varphi _{i}$, $\theta _{i}$ удовлетворяет соотношению $\partial _{x} \theta _{i} =-\Pi _{i}$. Первая строка в \eqref{eq:10:5.11} соответствует системе отдельных цепочек и имеет форму гамильтониана стандартной модели синус-Гордона. Первый член в \eqref{eq:10:5.11} описывает свободную бозе-систему, а второй соответствует взаимодействию бозонов вдоль цепочек, возникающего из-за рассеяния с процессом переброса («umklapp» рассеяния) фермионов, осуществляющих преобразование Йордана—Вигнера; последний вклад является маргинальным с РГ точки зрения и дает логарифмические поправки к термодинамическим величинам~\cite{10:037,10:092,10:093,10:094,10:095,10:096}. Численные оценки (см.~\cite{10:037,10:092}) приводят к значению взаимодействия $g_{\text{u}} /(2\pi )\simeq 0.25$. Вторая строка в \eqref{eq:10:5.11} описывает взаимодействие между цепочками.

\subsection{Приближение межцепочечного среднего поля для бозонизированного гамильтониана и~поправки первого порядка по~$1/z_{\bot }$}
\label{sec:10:5.4}

Простейший способ рассмотрения межцепочечного обменного взаимодействия~— так называемое межцепочечное приближение среднего поля~\cite{10:037}. Расцепляя член взаимодействия согласно
\begin{equation} \label{eq:10:5.12} 
\cos (\beta \varphi _{i} )\cos (\beta \varphi _{i+\delta _{\bot }} )\to 2\langle \cos (\beta \varphi _{i+\delta _{\bot }} )\rangle \cos (\beta \varphi _{i} ),  
\end{equation} 
находим
\begin{equation} \label{eq:10:5.13} 
\mathscr{H}_{\text{MF}} =\frac{v}{2} \sum _{i} \int dx \left[\Pi _{i}^{2} +(\partial _{x} \varphi _{i} )^{2} \right]+g_{\text{u}} \sum _{i} \int dx \cos 2\beta \varphi _{i} -\frac{\lambda }{\pi } h_{\text{MF}} \sum _{i} \int dx \cos (\beta \varphi _{i} ) 
\end{equation} 
где 
\begin{equation} \label{eq:10:5.14} 
h_{\text{MF}} =z_{\bot } J^{\prime } \frac{\lambda }{\pi} \langle \cos (\beta \varphi _{i} )\rangle , 
\end{equation} 
$z_{\bot }$~— число ближайших соседей в поперечном к цепочке направлении ($z_{\bot } =4$ для тетрагональной решетки). Приближение \eqref{eq:10:5.12} дает возможность свести проблему многих цепочек к проблеме одной цепочки в эффективном подрешеточном магнитном поле. Вводя функцию 
\begin{equation} \label{eq:10:5.15} 
B(h;T)=\frac{\lambda }{\pi } \langle \cos (\beta \varphi _{i} )\rangle _{h} ,  
\end{equation} 
вычисляемую в присутствии магнитного поля (последний член в~\eqref{eq:10:5.13}), получаем самосогласованное уравнение для подрешеточной намагниченности $\bar{S}$ 
\begin{equation} \label{eq:10:5.16} 
\bar{S}_{\text{MF}} =B(z_{\bot } J^{\prime } \bar{S}_{\text{MF}} ;T).  
\end{equation} 
Несмотря на то, что гамильтониан $\mathscr{H}_{\text{MF}}$ имеет одноцепочечную форму, вычисление функции $B(h;T)$ (являющейся аналогом функции Бриллюэна в обычной теории среднего поля гейзенберговских магнетиков) при произвольных температурах~— достаточно сложная задача. Согласно размерной оценке, $B(h;T)=h^{1/3} f(h^{2/3} /T)$ с~некоторой функцией $f(x)$,  $f(x)\sim x$ при $x\to 0$ и $f(\infty )=\const $. Для $g_{\text{u}} =0$ (в~этом случае имеем стандартную модель синус-Гордона или, что эквивалентно, массивную модель Тирринга) $B(h;T)$ была определена с помощью Бете-анзаца~\cite{10:097}. При~$h\to 0$
\begin{equation} \label{eq:10:5.17} 
B(h,T)=h\chi _{0} (T),  
\end{equation} 
где $\chi _{0} (T)$~— подрешеточная восприимчивость системы в отсутствии поля $h$~\cite{10:037,10:096},
\begin{equation} \label{eq:10:5.18} 
\chi _{0} (T)=\frac{\tilde{\chi }_{0}}{T} L\left(\frac{\Lambda J}{T} \right),\quad 
\tilde{\chi }_{0} =\frac{\Gamma ^{2} (1/4)}{4\Gamma ^{2} (3/4)} \simeq 2.1884, 
\end{equation} 
\begin{equation} \label{eq:10:5.19} 
L(\Lambda J/T)=C\left[\ln \frac{\Lambda J}{T} +\frac{1}{2} \ln \ln \frac{\Lambda J}{T} +O(1)\right]^{1/2} .  
\end{equation} 
Константы $C$ и $\Lambda $ могут быть определены на основании численных расчетов~\cite{10:098}: $C\simeq 0.137$, $\Lambda \simeq 5.8$. 

Результат \eqref{eq:10:5.17} дает возможность вычислить значение $T_{\text{N}}$ в теории среднего поля, поскольку $h_{\text{MF}} \to 0$ при $T\to T_{\text{N}}$. Уравнение для температуры Нееля имеет вид~\cite{10:037} 
\begin{equation} \label{eq:10:5.20} 
T_{\text{N}}^{MF} =z_{\bot } J^{\prime } \tilde{\chi }_{0} L(\Lambda J/T_{\text{N}}^{MF} ).  
\end{equation} 
Таким образом, согласно межцепочечной теории среднего поля $T_{\text{N}} \propto |J^{\prime }|$; подрешеточная намагниченность основного состояния $\bar{S}_{0} \propto \sqrt{|J^{\prime }|/J}$ также зависит степенным образом от $J^{\prime }$, что означает возникновение дальнего порядка при произвольно малых $|J^{\prime }|$. Эти результаты противоречат стандартной теории спиновых волн, которая не делает различия между целыми и полуцелыми значениями спинов и предсказывает конечное критическое значение $J_{\text{c}}^{\prime } \sim Je^{-\pi S}$~\cite{10:031,10:054}, так что при $|J^{\prime } |<J_{\text{c}}^{\prime }$ подрешеточная намагниченность $\bar{S}_{0}$ исчезает и 
\begin{equation} \label{eq:10:5.21} 
\bar{S}_{0} \propto \ln |J^{\prime }/J^{\prime }_{\text{c}} |,\quad 
T_{\text{N}} \propto \bar{S}_{0} \sqrt{|J^{\prime }|} 
\end{equation} 
при $|J^{\prime }|>J^{\prime }_{\text{c}}$. Указанное противоречие было разрешено с помощью метода ренормгруппы~\cite{10:034,10:035,10:036}, показавшего, что на масштабе обратной длины $\mu \gg J_{\text{c}}^{\prime } /J$ стандартная спин-волновая теория действительно применима, причем перенормировочный фактор намагниченности $Z_{\mu }^{-1/2} \propto \ln \mu $. С~другой стороны, для полуцелых спинов при $\mu \ll J_{\text{c}}^{\prime } /J$имеет место зависимость $Z_{\mu }^{-1/2} \propto \mu ^{1/2}$~\cite{10:034,10:035}, означающая справедливость результатов теории межцепочечного среднего поля при $|J^{\prime } |\ll J_{\text{c}}^{\prime }$. 

В то~же время численные значения температуры Нееля в межцепочечной теории среднего поля оказываются сильно завышенными по сравнению с экспериментальными данными, поскольку эта теория не принимает во внимание эффекты корреляций между спинами, расположенными на разных цепочках. В~частности, значение температуры Нееля \eqref{eq:10:5.20} не чувствительно к пространственной размерности системы, хотя в случае $d=1+1$ (оба измерения являются пространственными, второе соответствует направлению, поперечному по отношению к цепочкам) должно быть $T_{\text{N}} =0$; для случая $d=1+2$ значения $T_{\text{N}}$ оказываются слишком высокими по сравнению с экспериментальными данными. 

Корреляции между положениями спинов на разных цепочках выражаются в наличии коллективных возбуждений, вносящих вклад в термодинамические свойства. При этом ситуация в межцепочечной теории среднего поля аналогична недостаткам теории Стонера для зонных магнетиков, которая пренебрегает вкладом коллективных возбуждений, позже учтенных в теории Мории~\cite{10:024}. Как и в теории Мории, коллективные возбуждения в модели Гейзенберга могут быть рассмотрены в рамках приближения случайных фаз (ПСФ), в котором они определяются полюсами спиновых восприимчивостей~\cite{10:037,10:093}
\begin{equation} \label{eq:10:5.22} 
\chi ^{+-} (q_{z} ,\omega )=\frac{\chi _{0}^{+-} (q_{z} ,\omega )}{1-J^{\prime } (q_{x} ,q_{y} )\chi _{0}^{+-} (q_{z} ,\omega )/2} ,  
\end{equation} 
\begin{equation} \label{eq:10:5.23} 
\chi ^{zz} (q_{z} ,\omega )=\frac{\chi _{0}^{zz} (q_{z} ,\omega )}{1-J^{\prime } (q_{x} ,q_{y} )\chi _{0}^{zz} (q_{z} ,\omega )} ,  
\end{equation} 
где для тетрагональной решетки  
\begin{equation} \label{eq:10:5.24} 
J^{\prime } (q_{x} ,q_{y} )=2J^{\prime } (\cos q_{x} +\cos q_{y} ), 
\end{equation} 
$\chi _{0} (q,\omega )$~— динамическая подрешеточная восприимчивость в модели~\eqref{eq:10:5.13}. При $h\to 0$ восприимчивость $\chi _{0} (q,\omega )$ также определяется простыми аналитическими выражениями~\cite{10:095,10:096}: 
\[
\chi _{0} (q_{z} ,\omega )=\frac{1}{T} L\left(\frac{\Lambda }{T} \right)\tilde{\chi }_{0} (q_{z} /T,\omega /T),
\] 
\begin{equation} \label{eq:10:5.25} 
\tilde{\chi }_{0} (k,\nu )=\frac{1}{4} \frac{\Gamma (1/4+ik_{+} )\Gamma (1/4+ik_{-} )}{\Gamma (3/4+ik_{+} )\Gamma (3/4+ik_{-} )} ,\quad k_{\pm } =\frac{\nu \pm k}{4\pi } .  
\end{equation} 
При этом $\chi _{0} (0,0)=\chi _{0} (T)$. 

Чтобы определить поправки к межцепочечной теории среднего поля, связанные с вкладом коллективных возбуждений, можно использовать $1/z_{\bot }$-разложение ($z_{\bot }$~— число ближайших соседей в направлениях поперечных к цепочкам)~\cite{10:099}. Этот подход подобен $1/z$ разложению (или разложению по обратному радиусу взаимодействия), использовавшемуся много лет назад для улучшения стандартной теории среднего поля гейзенберговских магнетиков~\cite{10:100,10:101}; он позволяет определить температуру Нееля квазиодномерных систем с большей точностью, чем в межцепочечном приближении среднего поля. Для намагниченности подрешетки при этом получается результат 
\begin{equation} \label{eq:10:5.26} 
\bar{S}=\frac{1}{T} h_{\text{MF}} \tilde{\chi }_{0} L\left(\frac{\Lambda }{T} \right)\left\{1+\frac{\pi ^{2}}{2T\tilde{\chi }_{0}} L\left(\frac{\Lambda }{T} \right)\int d^{2} \mathrm{r} V(\mathrm{r})\left[\frac{1}{8} F(\mathrm{r})+\frac{1}{2} G(\mathrm{r})\right] \right\}, 
\end{equation} 
где 
\begin{equation} \label{eq:10:5.27}
V^{+-,zz} (\mathrm{x})=T\sum _{i\omega _{n}} \int \limits_{-\pi }^{\pi } \frac{dq_{z}}{2\pi } \sum _{q_{x} ,q_{y}} \frac{J^{\prime } (q_{x} ,q_{y} )\exp (iq_{z} x-i\omega _{n} \tau )}{1+\delta -J^{\prime } (q_{x} ,q_{y} )\chi _{0}^{+-,zz} (q_{z} ,\omega )} ,
\end{equation}
\begin{equation} \label{eq:10:5.28} 
\tilde{\chi }_{0} =\frac{\pi }{2} \int d^{2} \mathrm{z}\frac{1}{|\tilde{\varsigma }(\mathrm{z})|} \simeq 2.1184 
\end{equation} 
и функции $F(\mathrm{r})$, $G(\mathrm{r})$ определены в~\cite{10:099}. С~использованием связи между средним полем и подрешеточной намагниченностью \eqref{eq:10:5.14} после собирания всех поправок в знаменатель результат для температуры Нееля в первом порядке по $1/z_{\bot }$ принимает вид
\begin{equation} \label{eq:10:5.29} 
T_{\text{N}} =kJ^{\prime } z_{\bot } \tilde{\chi }_{0} L(\Lambda /T_{\text{N}} ).  
\end{equation} 
Результат \eqref{eq:10:5.29} отличается от результата тории среднего поля \eqref{eq:10:5.20} множителем~$k$, зависящем от структуры решетки в направлении, перпендикулярном к цепочкам. Численный расчет для случая $d=1+2$ на тетрагональной решетке приводит к значению $k\simeq 0.70$. Таким образом, уменьшение $T_{\text{N}}$ благодаря флуктуационным эффектам составляет $25\%$ его средне-полевого значения, что находится в хорошем согласии с результатами численного анализа~\cite{10:102}. Для $d=1+1$ имеем $k=0$, так что $T_{\text{N}} =0$. 

Аналогичные вычисления подрешеточной намагниченности основного состояния приводят к результату~\cite{10:099} 
\[
\bar{S}=\bar{S}_{0} -\frac{\Delta }{4\pi } \frac{\partial \Delta }{\partial h} I,
\] 
\begin{equation} \label{eq:10:5.30} 
I=\sum _{\mathbf{q}} \left[(1-\Gamma _{\mathbf{q}}^{\prime } /2)\ln \frac{1}{1-\Gamma _{\mathbf{q}}^{\prime }} +(3-Z^{\prime } \Gamma _{\mathbf{q}}^{\prime } /2Z)\ln \frac{1}{1-Z^{\prime } \Gamma _{\mathbf{q}}^{\prime } /(3Z)} \right],  
\end{equation} 
где $\Gamma _{\mathbf{q}}^{\prime } =\cos q$ для $d=1+1$ и $\Gamma _{\mathbf{q}}^{\prime } =(\cos q_{x} +\cos q_{y} )/2$ для $d=1+2$. Для тетрагональной решетки
\begin{equation} \label{eq:10:5.31} 
\bar{S}_{0} =(0.677-I)h_{\text{MF}}^{1/3} .  
\end{equation} 
Последний член в скобках в \eqref{eq:10:5.31} представляет собой $1/z_{\bot }$-поправку к намагниченности основного состояния, 
\begin{equation} \label{eq:10:5.32} 
I=\begin{cases}
0.011, & d=1+2,\\
0.060,\quad & d=1+1.
\end{cases}
\end{equation} 
Из \eqref{eq:10:5.31} следует, что намагниченность основного состояния уменьшается почти на $10\%$ по сравнению с ее значением в теории среднего поля для $d=1+1$ и только на $2\%$ в случае $d=1+2$. Таким образом, флуктуационные поправки для подрешеточной намагниченности основного состояния гораздо менее важны, чем для температуры Нееля, и в трехмерном случае ими можно пренебречь.

\subsection{Сравнение с~экспериментальными данными}
\label{sec:10:5.5}

Рассмотрим применение полученных результатов к описанию  экспериментальных данных магнитных квазиодномерных систем. Наиболее изученным квазиодномерным соединением является KCuF$_{3}$, имеющее спин $S=1/2$. Эксперименты нейтронного рассеяния~\cite{10:010} приводят к параметру магнитного обмена вдоль цепочек для этого соединения $J=406$~К и намагниченности основного состояния $\bar{S}_{0} /S=0.25$. Как обсуждается в~\cite{10:037}, это значение $\bar{S}_{0}$ соответствует $J^{\prime }/J=0.047$, так, что $J^{\prime } =19.1$~К. Межцепочечное приближение среднего поля \eqref{eq:10:5.20} приводит к значению $T_{\text{N}} =47$~К при этих параметрах, что несколько выше экспериментального результата $T_{\text{N}} =39$~К~\cite{10:010}. В~то~же время, результат $1/z_{\bot }$-разложения \eqref{eq:10:5.29} $T_{\text{N}} =37.7$~К находится гораздо ближе к экспериментальному значению. Таким образом, рассматриваемый подход слегка переоценивает флуктуационные эффекты, но значительно улучшает межцепочечное приближение среднего поля. Вклад двойного логарифмического члена в \eqref{eq:10:5.19} составляет приблизительно $5\%$ и улучшает согласие с экспериментальными данными. 

Другое соединение с $S=1/2$, широко обсуждаемое в литературе,~— Sr$_{2}$CuO$_{3}$~— имеет следующие параметры~\cite{10:011,10:012}: $J=2600$~К, $T_{\text{N}} =5$~К. Надежные экспериментальные данные для $J^{\prime }$ отсутствуют, но, используя \eqref{eq:10:5.29} и экспериментальное значение $T_{\text{N}}$, находим $J^{\prime } =1.85$~К. Тогда из  \eqref{eq:10:5.31} следует $\bar{S}_{0} /S=0.042$, что находится в согласии с экспериментальными данными ($\bar{S}_{0} /S \lesssim 0.05$). 

Для Ca$_{2}$CuO$_{3}$ экспериментальные параметры имеют следующие значения~\cite{10:011,10:012}: $S=1/2$, $J=2600$~К и $T_{\text{N}} =11$~К. Из них находим $J^{\prime } =4.3$~К и $\bar{S}_{0} /S=0.062$, что снова находится в хорошем согласии с экспериментальными данными~\cite{10:012}, которые дают $\bar{S}_{0}$(Ca$_{2}$CuO$_{3}$)/$\bar{S}_{0}$(Sr$_{2}$CuO$_{3}$)${}=1.5\pm 0.1$. Таким образом, результат \eqref{eq:10:5.29} достаточен для количественного описания реальных квазиодномерных магнитных систем.

\section*{Заключение}
\addcontentsline{toc}{section}{Заключение}

Квазиодномерные и слоистые магнетики представляют собой пример систем с сильными флуктуациями и нетривиальным поведением термодинамических и магнитных свойств. Исследование этих систем~— весьма нетривиальная проблема с точки зрения теоретической физики. Обычная спин-волновая теория (и даже ее усовершенствованный самосогласованный вариант~— ССВТ), хотя и приводит к правильному результату для температуры перехода $T_{\text{M}}$ в ведущем логарифмическом приближении, оказывается количественно применимой лишь при температурах $T\ll T_{\text{M}}$. В~области более высоких температур необходим учет динамического взаимодействия спиновых волн, выходящий за рамки низшего (борновского) приближения, а также существенно не спин-волновых (в частности, продольных) возбуждений. 

Проблема описания термодинамических свойств квазиодномерных и слоистых магнетиков получила существенное развитие в рамках теоретико-полевых методов, примененных к широко распространенной модели магнетизма этих систем~— модели Гейзенберга. Использование этих подходов позволяет получить простые аналитические результаты для температурной зависимости намагниченности и величины $T_{\text{M}}$, которые могут быть использованы при практической обработке экспериментальных данных. В~квазидвумерных магнетиках в широкой температурной области ниже $T_{\text{M}}$ спин-волновая картина спектра возбуждений является адекватной и взаимодействие спиновых волн приводит к появлению поправочных слагаемых в выражениях для намагниченности и обратной температуры Нееля $1/T_{\text{M}}$, значительно улучшающих согласие с экспериментальными данными. Узкая критическая область вблизи $T_{\text{M}}$ может быть описана с учетом неспинволновых возбуждений, в том числе в рамках $1/N$-разложения. В~квазиодномерных магнетиках переход к бозевским (не спин-волновым) возбуждениям позволяет построить систематическое разложение по обратному координационному числу решетки в направлениях, перпендикулярных к цепочкам.

Таким образом, с теоретической точки зрения в последнее время достигнуто хорошее понимание физической картины спектра и свойств низкоразмерных магнетиков в широком интервале температур. Оно дает основу для количественного описания свойств реальных систем, и мы ставили одной из своих задач привлечь внимание экспериментаторов к этому факту. В~то~же время при детальном анализе магнетизма конкретных соединений необходим учет дипольного взаимодействия, релятивистских взаимодействий типа Дзялошинского—Мории и~т.~д. Несмотря на то, что уже имеются первые попытки описания систем с такими взаимодействиями в рамках самосогласованного спин-волнового и теоретико-полевого подходов~\cite{10:103}, они ждут своего дальнейшего развития. С~другой стороны, широко исследуемые в последнее время комплексные соединения со сложной кристаллической структурой, а также системы типа пленок и мультислоев, рассматривавшиеся ранее в рамках спин-волновой теории~\cite{10:104}, требуют более конкретного изучения в рамках описанных под\-ходов.\looseness=1

Близкие проблемы возникают при описании систем, имеющих фрустрированные магнитные структуры~— на двумерной квадратной решетке с учетом обменных взаимодействий между следующими за ближайшими соседями~\cite{10:105,10:106,10:107,10:108,10:109}, двумерной треугольной решетке~\cite{10:110,10:111,10:112,10:113,10:114,10:115},  решетках Кагоме, пирохлора, \cite{10:116,10:117} и~т.~д. Наличие спиновых фрустраций в таких системах приводит, как и в низкоразмерных соединениях, к подавлению дальнего магнитного порядка (при сохранении ближнего) и, следовательно, к очень нетривиальным термодинамическим свойствам. Фрустрированные системы также рассматривались в рамках спин-волновых теорий~\cite{10:118,10:119,10:120,10:121,10:122,10:123}.

Еще одна проблема, актуальная, например, в связи с высокотемпературной сверхпроводимостью и не затронутая нами,~— взаимодействие носителей тока с магнитными моментами. Специфика низкоразмерных систем (сильный ближний магнитный порядок) приводит к соответствующим особенностям электронного спектра~\cite{10:124,10:125}. Сильное электрон-электронное взаимодействие в этих условиях является дополнительным фактором, приводящим к формированию некогерентных электронных состояний и возможности перехода металл-изолятор. В~связи с этим сейчас ведется интенсивное теоретическое и экспериментальное исследование проводящих низкоразмерных систем, находящихся вблизи такого перехода~\cite{10:126,10:127,10:128}. Оно требует развитие существенно новых подходов, в которых, однако, могут быть использованы теоретические методы описания подсистемы локализованных моментов.\looseness=-1

Обсуждаемые низкоразмерные системы характеризуются малыми значениями точки магнитного перехода и в ряде случаев (особенно в квазиодномерной ситуации) малым моментом основного состояния, что сближает их со слабыми зонными магнетиками. Эта аналогия является достаточно глубокой и раскрывается в теоретико-полевых подходах, описывающих спектр возбуждений системы с помощью нестандартных представлений (например, метод бозонизации, андерсоновские спиноны и~пр.). В~последнее время рассматриваются подходы, которые позволяют построить единое описание низкоразмерных и фрустрированных гейзенберговских систем и решеток Кондо, также обладающих малыми моментами~\cite{10:129}.

\end{document}